\documentclass{aa}

\usepackage{graphicx}
\usepackage{gensymb}
\usepackage{hyperref}
\hypersetup{colorlinks=true,linkcolor=blue,citecolor=blue,filecolor=blue,urlcolor=blue,}
\newcommand{\orcid}[1]{\protect\href{https://orcid.org/#1}{\protect\includegraphics[width=10pt]{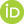}}}
\usepackage{txfonts}

\newcommand{\ha}{H$\alpha$\,} 
\newcommand{\EWha}{EW$_{H\alpha}$} 
\newcommand{\hb}{H$\beta$\,} 
\newcommand{\SBha}{$\Sigma_{H\alpha}$\,} 

\newcommand{\hii}{H\,{\footnotesize II}~\,}
\newcommand{\hei}{He\,{\footnotesize I}\,}
\newcommand{\nii}{[N\,{\footnotesize II}]\,}
\newcommand{\sii}{[S\,{\footnotesize II}]\,}

\newcommand{\oi}{[O\,{\footnotesize I}]\,}
\newcommand{\oii}{[O\,{\footnotesize II}]\,}
\newcommand{\oiii}{[O\,{\footnotesize III}]\,}
\defcitealias{2022A&A...659A..26B}{Bel22}
\defcitealias{2017MNRAS.466.3217Z}{Z17}
\defcitealias{2018MNRAS.474.3727L}{Lac18}
\begin{document}

   \title{Bidimensional
Exploration of the warm-Temperature Ionised gaS (BETIS)}

   \subtitle{I. Showcase sample and first results}

   \author{R. Gonz\'alez-D\'iaz
          \inst{1,2}\orcid{0000-0002-0911-5141}
          \and
          F. F. Rosales-Ortega
          \inst{2}\orcid{0000-0002-3642-9146}
          \and
          L. Galbany
          \inst{1,3}\orcid{0000-0002-1296-6887}
          \and
          J. P. Anderson
          \inst{4,5}\orcid{0000-0003-0227-3451}
          \and
          C. Jim\'enez-Palau
          \inst{1}\orcid{0000-0002-4374-0661}
          \and
          M. Kopsacheili
          \inst{1}\orcid{0000-0002-3563-819X}
          \and
          H. Kuncarayakti
          \inst{8,9}\orcid{0000-0002-1132-1366}
          \and
          J. D. Lyman
          \inst{6}\orcid{0000-0002-3464-0642}
          \and
          S. F. S\'anchez
          \inst{7}\orcid{0000-0001-6444-9307}
          }

   \institute{Institute of Space Sciences (ICE-CSIC), Campus UAB, Carrer de Can Magrans, s/n, E-08193 Barcelona, Spain.\\ 
              \email{Raul.GonzalezD@autonoma.cat}
         \and
         Instituto Nacional de Astrof\'isica, \'Optica y Electr\'onica (INAOE-CONAHCyT),
              Luis E. Erro 1, 72840 Tonantzintla, Puebla, M\'exico.
         \and
            Institut d’Estudis Espacials de Catalunya (IEEC), E-08034 Barcelona, Spain.
         \and
             European Southern Observatory, Alonso de Córdova 3107, Vitacura,
             Casilla 19001, Santiago, Chile.
         \and
             Millennium Institute of Astrophysics MAS, Nuncio Monseñor
             Sotero Sanz 100, Off. 104, Providencia, Santiago, Chile.
         \and
             Department of Physics, University of Warwick, Coventry CV4 7AL, UK.
         \and
          Instituto de Astronom\'ia, Universidad Nacional Aut\'onoma de M\'exico, A. P. 70-264, C.P. 04510, Ciudad de M\'exico, M\'exico.
         \and
          Finnish Centre for Astronomy with ESO (FINCA), FI-20014 University of Turku, Finland.
         \and
          Tuorla Observatory, Department of Physics and Astronomy, FI-20014 University of Turku, Finland.
        }

   \date{Received ... ; accepted ... }

 
\abstract{
We present the Bidimensional Exploration of the warm-Temperature Ionised gaS (BETIS) project, designed for the spatial and spectral study of the diffuse ionised gas (DIG) in a selection of nearby spiral galaxies observed with the MUSE integral-field spectrograph. Our primary objective is to investigate the various ionisation mechanisms at play within the DIG. We analysed the distribution of high- and low-ionisation species in the optical spectra of the sample on a spatially resolved basis.
We introduced a new methodology for spectroscopically defining the DIG, optimised for galaxies of different resolutions. Firstly, we employed an innovative adaptive binning technique on the observed datacube based on the spectroscopic signal-to-noise ratio (S/N) of the collisional \sii line to increase the S/N of the rest of the lines including \oiii, \oi, and \hei. Subsequently, we created a DIG mask by eliminating the emissions associated with both bright and faint \hii regions. We also examined the suitability of using \ha equivalent width (\EWha) as a proxy for defining the DIG and its associated ionisation regime. Notably, for \EWha$<3$\r{A}  -- the expected emission from hot low-mass evolved stars (HOLMES) – the measured value is contingent on the chosen population synthesis technique performed.
Our analysis of the showcase sample reveals a consistent cumulative DIG fraction across all galaxies in the sample, averaging around 40\%-70\%. The average radial distribution of the \nii/\ha, \sii/\ha, \oi/\ha, and \oiii/\hb ratios are enhanced in the DIG regimes (up to 0.2 dex). It follows similar trends between the DIG regime and the \hii regions, as well as the \ha surface brightness (\SBha), indicating a correlation between the ionisation of these species in both the DIG and the \hii regions. 
The DIG loci in typical diagnostic diagrams are found, in general, within the line ratios that correspond to photoionisation due to the star formation. There is a noticeable offset correspondent to ionisation due to fast shocks. However, an individual diagnosis performed for each galaxy reveals that all the DIG in these galaxies can be attributed to photoionisation from star formation. The offset is primarily due to the contribution of Seyfert galaxies in our sample, which is closely aligned with models of ionisation from fast shocks and galactic outflows, thus mimicking the DIG emission. Our results indicate that galaxies exhibiting active galactic nucleus (AGN) activity should be considered separately when conducting a general analysis of the DIG ionisation mechanisms, since this emission is indistinguishable from high-excitation DIG.
}

   \keywords{galaxies: ISM – galaxies: star formation – \hii regions – ISM: structure – ISM: general}

   \maketitle
%
\section{Introduction}

Understanding the relationship between stellar formation processes and the interstellar medium (ISM) is a key step in discerning the complexity in the evolutionary history of galaxies. One problem in this regard has been understanding the nature and importance of feedback processes in which massive stars deposit energy into the interstellar medium through photoionisation, stellar winds, and supernovae. This feedback mechanism affects the physical and dynamic state of the ISM and, therefore, it has an influence on the rate and distribution of stellar formation in galaxies \citep{2008IAUS..245...33C,2009ApJ...695..292C, 2014MNRAS.445..581H, 2016SAAS...43...85K, 2017MNRAS.466.1903G, 2017PhDT........59G}.

In this context, the existence of a warm and ionised component of the ISM that is ubiquitously distributed in our galaxy has been known for decades. This component of the Milky Way has been referred to as the warm ionised medium (WIM) \citep{1963AuJPh..16....1H,1968Natur.217..709H,1971PhDT.........1R, 1973ApJ...179..651R, 1974IAUS...60...51G,1985ApJ...294..256R, 1989ApJ...345..811R, 1989ApJ...337..761K, 2003ApJS..146..407F}. With the warm temperatures and low densities found in this medium ($0.6-1\cdot10^4$ K and $\sim10^{-1}$cm$^{-3}$ respectively, with an emission measurement of $\sim35$ cm$^{-6}$ pc), the WIM represents the 20\% of the ISM in volume and the 90\% of the ionising hydrogen in mass, being one of the most important components of our galaxy \citep{1971PhDT.........1R, 1971A&A....12..379M}. The intensity of the forbidden \sii $\lambda$6716 and \nii $\lambda$6584 lines is found to be higher with respect to the \ha intensity in the WIM in comparison with the typical intensities of the \hii regions; in addition, the measurement of these lines reveals that the ionisation state of the WIM and its temperature vary significantly along the galactic plane, increasing towards the galactic centre \citep{1985ApJ...294..256R,2005ApJ...630..925M}. Moreover, the filling fraction increases as the height with respect to the galactic plane increases, being around 0.1 at the midplane and reaching 0.3-0.4 at 1 kpc from the galactic plane \citep{1987ASSL..134...87K, 1991IAUS..144...67R, 2006AN....327...82B,2008PASA...25..184G}.

The first detection of this medium in other galaxies, now referred as diffuse ionised gas (DIG) in the extragalactic context, were carried out through narrowband \ha images of edge-on star-forming galaxies \citep{1990ApJ...352L...1R, 1990A&A...232L..15D}. This was followed up by studies carried by small samples (< 5) of face-on and edge-on galaxies using narrowband \ha and \sii images \citep{1996AJ....111.2265F,1997ApJ...483..666G, 2000A&A...363....9Z} and long-slit spectroscopy with limited spatial coverage \citep{1985ApJ...294..256R, 1997ApJ...491..114W,2001ApJ...560..207O, 2003ApJ...586..902H, RevModPhys.81.969}.

All these studies have shown that 30\%$-$50\% of the total \ha luminosity of their galaxies correspond to the DIG, being the $80\pm10\%$ of the projected area of the disks occupied by DIG \citep{2007ApJ...661..801O}, thereby proving the ubiquitousness and omnipresence of this component in the star-forming galaxies. Moreover, the spectroscopic studies shown that the emission of low-ionisation lines, such as \sii$\lambda\lambda6717,31$, \nii$\lambda6584,$ or \oi$\lambda6300$, are enhanced in DIG and WIM regimes in comparison with the emission of the \hii regions \citep{1997ApJ...491..114W}. Specifically, the ratios of those lines with respect the \ha emission in the DIG are larger than in \hii regions (found firstly in edge-on galaxies; \citealt{2001ApJ...560..207O}). The leakage of the Lyman continuum (Lyc) photons, produced by the OB stars within \hii regions, is commonly regarded as the primary source of DIG and WIM ionisation \citep{2009RvMP...81..969H, 2009ApJ...703.1159S, 2022A&A...659A..26B}. However, it has been observed that in some cases, the total energy output of these stars does not appear to provide the necessary power to ionise the DIG uniformly across galactic discs (\citealt{1996AJ....111.2265F, 1996AJ....112.2567F}). The substantial power demand, combined with the presence of high-ionisation species like \oiii (enhanced at more than 1 kpc above the galactic plane) and \hei \citep{1998ApJ...494L..99R,2004MNRAS.353.1126W, 2009RvMP...81..969H} underscores the insufficiency of Lyc photons leaking from \hii regions in explaining the existence of the DIG. The question of the ionisation source for this component continues to be a topic of ongoing debate, with numerous proposals suggesting alternative heating sources for the DIG. Some of these sources include photoelectric heating from interstellar dust particles or large molecules \citep{1992ApJ...400L..33R, 2001ApJS..134..263W}, fast shocks provided by Wolf-Rayet stars or supernovae \citep{1998ApJ...494L..99R, 2001ApJ...551...57C, 2005A&A...442..443H}, turbulent mixing layers and dissipation of turbulence \citep{1993ApJ...407...83S,1997ApJ...485..182M, 1998ApJ...494L..99R, 2009ApJ...695..552B}, cosmic ray heating \citep{2013ApJ...767...87W}, or microflares and magnetic field reconnections \citep{1992ApJ...384..502R, 1998MNRAS.296..165B}. 
Another alternative source of ionisation, as proposed by
\citet{2011MNRAS.415.2182F} and \citet{2011MNRAS.413.1687C}, is the photoionisation from hot low-mass evolved stars (HOLMES), that could provide an explanation for the high \oiii/\ha ratio observed in the extraplanar DIG \citep{1990ApJ...352L...1R}.

The extent to which HOLMES contribute to the ionising radiation responsible for the DIG remains uncertain. While they may play a role in ionising specific regions within the sparsely populated interstellar medium, their overall significance is not yet established. Initial calculations conducted by \citet{1974ApJ...190..109H} proposed that the ionising radiation emitted by these hot pre-white dwarf stars could have a substantial influence on the interstellar medium.

The exploration of the combination of Lyc photon leakage and HOLMES as potential primary sources of ionisation for the DIG has been extensively investigated using integral field spectrographs (IFS). The IFS offers simultaneous spatial and spectral information, enabling the examination of these ionisation mechanisms across a wide range of spatial resolutions and high spectral resolution. Some IFS surveys provides large samples of nearby galaxies at kiloparsec resolution such as the Calar Alto Legacy Integral Field spectroscopy Area survey (CALIFA; \citealt{2012A&A...538A...8S, 2013A&A...549A..87H}; 391 galaxies at $\sim$0.8 kpc resolution; \citealt{2018MNRAS.474.3727L}, hereafter \citetalias{2018MNRAS.474.3727L}) or the Mapping Nearby Galaxies at APO (MaNGA; \citealt{2015ApJ...798....7B}; 356 galaxies at 2 kpc resolution; \citealt{2017MNRAS.466.3217Z}, hereafter \citetalias{2017MNRAS.466.3217Z}).

\citetalias{2017MNRAS.466.3217Z} found an enhancement in the DIG of the ratios \sii/\ha, \nii/\ha, \oii/\ha, and \oi/\ha, increasing in function of the galactocentric radius, as well as a variable trend of the \oiii/\hb ratio in comparison with \hii regions. Besides, \citetalias{2018MNRAS.474.3727L} defines a criterion of DIG definition in function of the \ha equivalent width (\EWha), based on if the region if dominated by photoionisation by HOLMES, corresponding to a \EWha$<3$\r{A}. This would then be the primary regime in E-S0 galaxies, as the old stellar populations conform to these galaxies.

On the other hand, IFS studies based on the MUSE instrument (Multi Unit Spectroscopic Explorer; \citealt{2010SPIE.7735E..08B}) or wide-field spectroscopic coverage such as TYPHOON \citep{2022arXiv221106005G} offer datasets with a spatial resolution ranging from $\sim50$ pc to the resolution of MaNGA-like galaxies, which allows for a spatially resolved study of the DIG and \hii regions. Nevertheless, the studies of the DIG using resolutions of $\sim50$ pc have predominantly focused on the best galaxies within these surveys in terms of spatial resolution. These studies typically involve individual galaxies, such as M83 \citep{2019MNRAS.487...79P, 2022A&A...660A..77D, 2022A&A...666A..29D}, or the 19 galaxies encompassed by the  Physics at High Angular resolution in Nearby GalaxieS project (PHANGS; \citealt{2022A&A...659A.191E, 2022A&A...659A..26B}, hereafter \citetalias{2022A&A...659A..26B}), capable of resolving the ISM structure at $\sim 50$ pc of resolution. Studies conducted within the PHANGS galaxies reveals that photoionisation by HOLMES is more prevalent in central regions where \EWha$<3$ \r{A}. In these regions, approximately 2\% of the total \ha emission is powered by HOLMES. The remaining emission arises from photon leakage from \hii regions. The mean free path of ionised photons is estimated to be $<1.9$ kpc, based on a straightforward thin-slab model for photon propagation through the ISM \citep{2002A&A...386..801Z, 2009ApJ...703.1159S}. This model also predicts an increase in the \oiii/\hb ratio and the ionisation parameter, which contradicts the decrease in the ratio as \SBha increases. However, the contribution of the spectral hardness of HOLMES to the ionisation budget explains this discrepancy and the trend of increasing \sii/\ha, \nii/\ha, and \oi/\ha line ratios as \SBha decreases. 

In order to make progress in quantifying the escape fraction of ionising photons from \hii regions, it is imperative to reassess the importance of leaked radiation and HOLMES in contributing to the ionisation of the DIG in star-forming disk galaxies. However, previous studies aimed at this task have relied on large surveys with low spatial resolution or small samples with high spatial resolution. In most cases, these studies have employed methodologies that primarily rely on \ha emissions for the definition and exploration of the DIG.

In this paper, we introduce the Bidimensional Exploration of the warm-Temperature Ionised gaS (BETIS) project. The main goals of BETIS are the exploration of potential ionisation mechanisms, assessment of the DIG's influence on calculating chemical abundances and star formation rates, and examination of the correlations between the DIG and various factors such as galaxy morphology, star formation rate, neutral hydrogen abundance, and other physical parameters. For this first paper of the BETIS series, we present the methodology employed in this project, designed to be applicable to galaxies with varying linear resolutions and morphologies, based not only on \ha, but on other emission lines. In addition, we discuss the problematical aspects of performing a general and global study of the ionisation mechanisms of the DIG. We validate this methodology through testing on a representative sample of seven galaxies with diverse characteristics. The goal is to determine the implications of incorporating galaxies exhibiting different physical processes into the same diagnostic of the ionisation mechanisms of the DIG.

This paper is structured as follows. Section \ref{SEC:sample} provides a definition of the showcase sample selection from different MUSE observations. Section \ref{SEC:method} outlines the methodology for distinguishing between DIG emission and \hii regions emission. We also introduce a novel adaptive binning method that will be applied to the observed datacubes, designed to enhance the signal-to-noise ratios (S/Ns) of our data. This methodology is tested on a showcase sample in Sect. \ref{SEC:results}, showing preliminary results, as well as a discussion of the challenges encountered when using \EWha \, as a proxy for characterising the DIG. Finally,  Sect. \ref{SEC:conclusions} provides a overview of the key findings and outcomes presented in this paper. Additionally, we outline our plans for future research, specifically focusing on the forthcoming parts of the DIG analysis with a full BETIS sample.

The following notation is used throughout the paper: \nii$\equiv$\nii$\lambda6584$; \sii$\equiv$\sii$\lambda6717 +$\sii$\lambda6731$; \oiii$\equiv$\oiii$\lambda5007$; and \oi$\equiv$\oi$\lambda6300$. We adopt the standard $\Lambda$CDM cosmology with H$_0$ = 70 km/s/Mpc,\, $\Omega_{\lambda}$ = 0.7, \,$\Omega_M$ = 0.3.

\section{The BETIS showcase sample}\label{SEC:sample}

The task of characterising and understanding the DIG in terms of star formation requires spatially resolved information, so that we may discern the DIG emission from the \hii regions and explore the possible ionisation mechanisms. For this reason, the IFS surveys are the ideal data sets for studying the DIG, for instance, MaNGA \citepalias{2017MNRAS.466.3217Z} and CALIFA \citepalias{2018MNRAS.474.3727L}, since they bring the necessary optical coverage to undertake this challenge. However, the limited spatial resolution has been the main obstacle in such studies. In terms of resolution and spectral coverage, MUSE is one of the most suitable instruments to study the DIG at extragalactic level.

MUSE is an IFS located at the ESO-VLT 8.1m telescope at Cerro Paranal, Chile. The instrument brings 1 arcmin$^2$ of field of view (FoV), with a sampling of $0.2\times0.2$ arcsec, a spectral range of $4650-9300$ \r{A}, with a resolution of 1750 at 4650 \r{A} to 3750 at 9300 \r{A}, and a spectral sampling of 1.25 \r{A}. 
The data observed by this instrument has already been used to study the DIG (PHANGS-MUSE; \citetalias{2022A&A...659A..26B}; \citealt{2022A&A...660A..77D, 2022A&A...666A..29D, 2023A&A...672A.148C}) at a spatial resolution of $\sim50$ pc, but constrained to only $<20$ objects. 

In order to develop a methodology to study the DIG at different physical resolutions (from PHANGS-like to CALIFA-like) and present initial findings of this analysis, a representative subset of seven galaxies of different morphologies, characteristics and resolutions has been selected from different MUSE projects (which will be included in the full BETIS sample for the forthcoming papers). This showcase sample has been carefully chosen to provide insight into the broader study and serves as a demonstration of the methodology employed. In addition, this sample will also serve to prove the impact of using galaxies of different characteristics and physics involved in the same analysis of the DIG.

\noindent
The showcase sample was selected from the All-weather MUse Supernova Integral-field of Nearby Galaxies (AMUSING;  \citealt{2016MNRAS.455.4087G}) and AMUSING+ (Galbany et al. in prep.) project samples.
These surveys are ongoing projects aimed at studying the environments of supernovae by means of the analysis of a large number of nearby supernova host galaxies
$(0.005 < z < 0.1$). The AMUSING survey\footnote{The AMUSING survey characterisation is available online: \href{https://amusing-muse.github.io/sample/}{https://amusing-muse.github.io/sample/}} currently comprises 571 supernova hosts observed during the semesters P95-P104 (April 2015-March 2020) and is composed by a wide variety of galaxy types with the common characteristic of having hosted a known supernova. The AMUSING+ sample is aimed to increase the number of supernova host, adding 143 objects to the previous AMUSING sample. The criteria followed for selecting the showcase sample galaxies from the AMUSING and AMUSING+ samples are as follows.

    First, to test the methodology for different linear resolutions, we chose galaxies with resolutions between  the full width at half
maximum for PHANGS and CALIFA  (41 pc $\lesssim$ FWHM $\lesssim$ 1 kpc). 
    Next, we aimed to analyse the radial behaviour of the DIG. However, since the physical properties of \hii regions can vary between the inner and outer parts of galaxies  \citep{2019A&A...631A..23R}, to avoid limiting ours study to the inner regions, we assumed that at least 0.5R$_{25}$\footnote{R$_{25}$ defined as the isophote at the blue brightness of 25 mag/arcsec$^2$.} of the galaxy must be in the datacube FoV if the resolution is high (FWHM $\lesssim$ 100 pc).
    Then, to test the methodology for different galaxy morphologies, we assumed every galaxy must have a different Hubble type, while also exhibiting star-forming regions.
    Finally, we did not consider galaxies with low depth or data-quality. In particular, if the average spectroscopic S/N of the \ha line in the DIG regions was lower than 3, the galaxy was disregarded.

\begin{figure*}[ht!]
    \includegraphics[width=\textwidth]{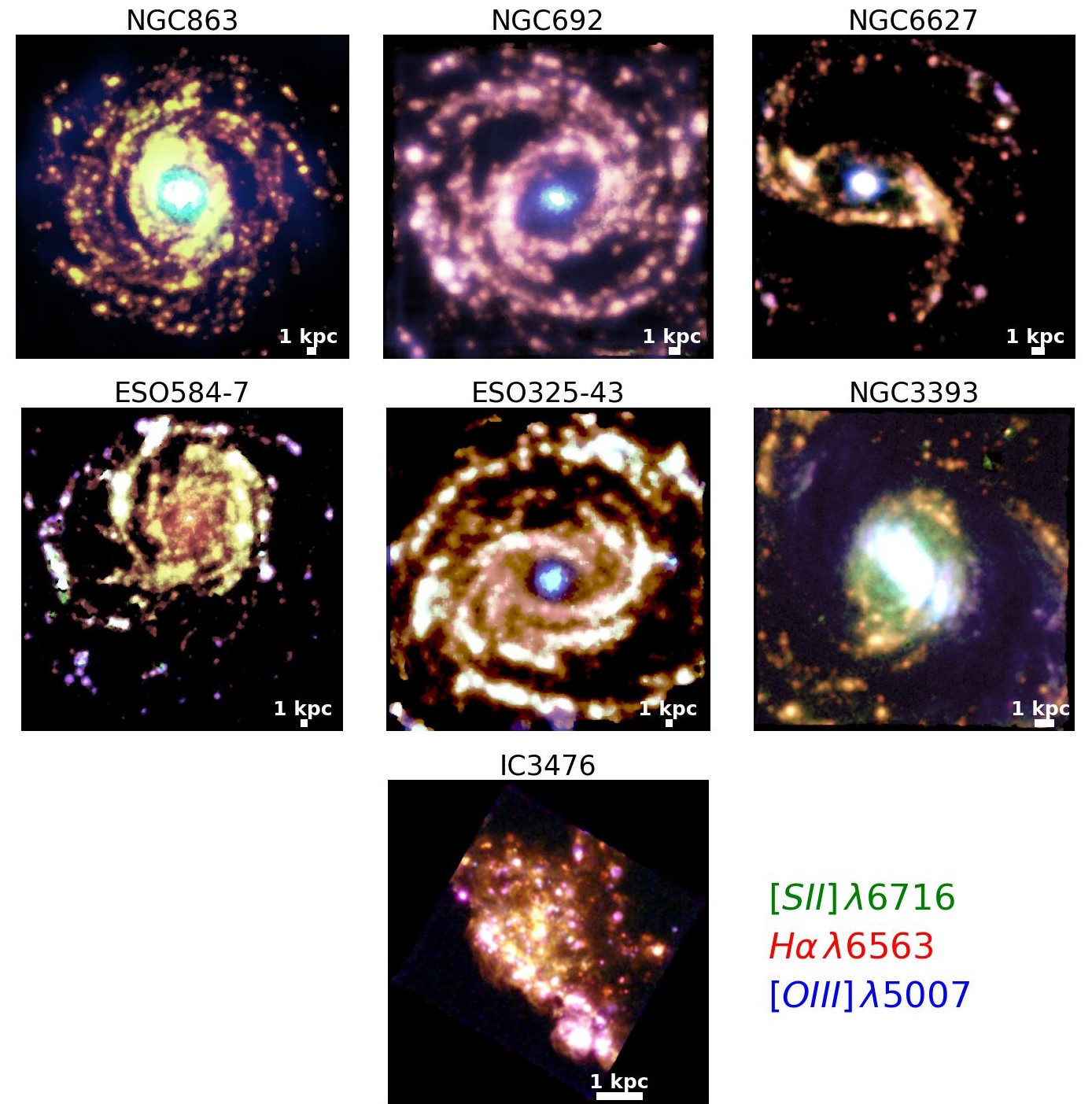}
    \caption{RGB synthetic images of the galaxies selected for our sample. The images are constructed as a composition of the \sii$\lambda6716$ (green), \ha (red), and \oiii$\lambda5007$ (blue) spaxel-by-spaxel emission maps obtained following the methodology explained in Sect. 3.2.}
    \label{fig:sub_collague}
\end{figure*}

Seven galaxies were chosen based on the previous criteria, belonging to the ESO periods P95, P98, P99, P101, and P103. The respective datacubes were already reduced by the AMUSING collaboration using the ESO reduction pipeline \citep{2016MNRAS.455.4087G, 2020A&A...641A..28W}. Table \ref{tab:TableSample} shows the general characteristics of the BETIS showcase sample and Table \ref{tab:obs} shows the characteristics of the observations. The FWHM measured of the galaxies were derived assuming the seeing reported by the telescope via the \texttt{TEL.IA.FWHMOBS} parameter. We caution that this value corresponds to the PSF measured from the differential image motion monitor (DIMM) at Paranal and propagated to the data headers. This value may be unreliable, especially in cases of high wind, however, we can assume this value as the seeing since we are not working with the native resolution (see Sect. \ref{SEC:method}). The integrated stellar mass of the galaxies, obtained from a single stellar population (SSP) fitting with {\sc STARLIGHT} (see Sect. \ref{SEC:method}), spans between $3.31\cdot10^{10}M_{\odot}<M_{\star}<1.04\cdot10^{12}M_{\odot}$, while the integrated star formation rate (SFR) spans between $1.31M_{\odot}/yr$ < SFR < $4.92M_{\odot}/yr$. The SFR was obtained using the \citet{2012ARA&A..50..531K} relation: log(SFR) = log($L_{H\alpha}$) - log($C_{H\alpha}$), with $L_{H\alpha}$ as the \ha luminosity, corrected for interstellar extinction assuming the Cardelli extinction law, assuming R$_V$ = 3.1 \citep{1989ApJ...345..245C} and a Balmer decrement \ha/\hb = 2.85. Then, log($C_{H\alpha}$) is the conversion factor between SFR and $L_{H\alpha}$, corresponding to 41.27 \citep{2012ARA&A..50..531K}. Figure \ref{fig:sub_collague} shows false-colour images of the sample constructed as a composition of the \sii (green), \ha (red), and \oiii (blue) emission line maps obtained following the methodology described in Sect. \ref{subSEC:fitting}, but on a spaxel-by-spaxel basis. The native spatial resolutions vary from 49 pc (IC3476, i.e. PHANGS-like) to 960 pc (ESO325-43, CALIFA-like), with a median resolution of 395 pc. The sample includes three known active galactic nuclei (AGNs): NGC863 and NGC3393 are Seyfert 2 \citep{1977ARA&A..15...69W, 1988SoSAO..55....5L}, while NGC692  is a low-luminosity AGN (LLAGN; \citealp{2019ApJ...872..134Z}). We deliberately incorporated these galaxies presenting AGN emission. Although the AGN emission of the centre of the galaxy are masked during the methodology and analysis of the results, our aim is to investigate the impact of including galaxies exhibiting different physical processes via a global analysis of the ionisation mechanisms of the DIG in galaxies.

\begin{table*}[ht!]
\centering
\resizebox{\textwidth}{!}{%
\begin{tabular}{llrrrrrrrrrrr}
\hline \hline
\textbf{Galaxy} & \textbf{Morphology} & \textbf{RA} & \textbf{DEC} & \textbf{D} & \textbf{z} & \textbf{FWHM} & \textbf{Incl} & \textbf{PA} & \textbf{spx size} & \textbf{log M$_{\star}$} & \textbf{log SFR} \\
 & & (J2000) & (J2000) & (kpc) & & (pc) & (°) & (°) & (pc) & (M$_{\odot}$) & (M$_{\odot}$/yr) \\ \hline
NGC863 & SA(s)a & 02h14m & -00d46m & 34 & 0.02639 & 591.98 & 22.40 & 175.6 & 109 & 12.02 & 0.37 \\
NGC3393 & (R')SB(rs)a & 10h48m & -25d09m & 69 & 0.01246 & 253.56 & 25.74 & 12.8 & 51 & 10.96 & 0.19 \\
NGC6627 & (R')SB(s)b & 18h22m & 15d41m & 28 & 0.01771 & 297.82 & 25.12 & 70.0 & 73 & 10.94 & 0.12 \\
NGC692 & (R')SB(r)bc & 01h48m & -48d38m & 96 & 0.02115 & 395.21 & 30.30 & 84.4 & 87 & 11.33 & 0.45 \\
ESO584-7 & Sc & 16h12m & -21d37m & 65 & 0.03171 & 618.88 & 45.12 & 148.0 & 131 & 10.52 & 0.40 \\
ESO325-43 & SAB(s)c pec & 13h59m & -37d51m & 107 & 0.03503 & 960.03 & 38.37 & 115.0 & 145 & 10.56 & 0.14 \\
IC3476 & IB(s)m & 12h32m & 14d03m & 10 & -0.00063 & 49.60 & 33.99 & 30.0 & 16 & 11.03 & 0.69 \\ \hline
\end{tabular}%
}
\caption{General characteristics of the BETIS showcase sample, in order of morphological type. The columns represents, from left to right: The designation of the galaxy, the morphological Hubble-De Vaculeurs type, the RA and DEC in the J2000 epoch of the centre of the galaxies, restricted from the Paranal observatory (DEC $<25$\degree), the physical diameter in kpc, the redshift, the PSF FWHM in pc, inclination with respect to the line of sight (in deg), the position angle (in deg) and the physical size of the spaxel, in parsec/spaxel, the log of the integrated stellar mass in solar masses, and the log of the star formation rate in solar masses per year. The RA, DEC, the diameter and the redshift are obtained from NED. The morphological type and position angle, from Hyperleda. M$_{\star}$ was obtained from \citet{2020AJ....159..167L}, except NGC3393 and NGC6627 that was obtained from SSP fitting. The SFR was obtained as explained in Sect. \ref{SEC:sample}.}
\label{tab:TableSample}
\end{table*}


\begin{table*}[ht!]
\centering
\resizebox{\textwidth}{!}{%
\begin{tabular}{lccccc}
\hline
\hline
\textbf{Galaxy} & \textbf{project} & \textbf{PI} & \textbf{ESO project ID} & \textbf{exp. time (s)} & \textbf{seeing (arcsec)}\\ \hline
NGC 0863 & \begin{tabular}[c]{@{}c@{}}The dynamics of the AGN\\ fuelling reservoir in MRK 590\end{tabular} & Sandra Raimundo & 099.B-0294 & 8x1100 & 0.901 \\
NGC 3393 & The MUSE atlas of disks (MAD) & C.M. Carollo & 098.B-0551 & 4x900 & 0.789 \\
NGC 6627 & AMUSING survey IX & J. Anderson & 103.D-0440 & 4x601 & 0.839 \\
NGC 0692 & AMUSING survey VII & Hanindyo Kuncarayakti & 101.D-0748 & 4x599 & 1.403 \\
ESO584-7 & AMUSING survey VII & Hanindyo Kuncarayakti & 101.D-0748 & 4x701 & 0.796 \\
ESO325-43 & AMUSING survey V & L. Galbany & 099.D-0022 & 8x701 & 1.285 \\
IC3476 & \begin{tabular}[c]{@{}c@{}}MUSE study of nearby CC-SNe host \\ environments and parent stellar populations \end{tabular} & Hanindyo Kuncarayakti & 095.D-0172 & 4x450 & 0.742 \\
\hline
\end{tabular}%
}
\caption{Characteristics of the observations. Second column is the project where the galaxy was observed. Last column is the measured seeing in arcesec (as the \texttt{TEL.IA.FWHMOBS} parameter in the header of the .fits files from the datacubes). All galaxies were observed following the AMUSING survey strategy: non-optimal weather at Paranal, that is, at any seeing and even a bright moon.}
\label{tab:obs}
\end{table*}

\section{Methodology}\label{SEC:method}

The study of extragalactic DIG has conventionally relied on analyses of high-resolution narrowband \ha and images of nearby galaxies ($z<0.01$, \citealt{1990ApJ...352L...1R, 1990A&A...232L..15D, 1996AJ....111.2265F,2000A&A...363....9Z}) or long-slit spectroscopy with limited spatial coverage \citep{1985ApJ...294..256R, 1997ApJ...491..114W,2001ApJ...560..207O, 2003ApJ...586..902H, RevModPhys.81.969}. In these studies, the methodology for subtracting the DIG from the galaxies and distinguish their emission from the \hii regions emission is typically based on using cut-offs in the surface brightness $(\Sigma_{H\alpha})$ of the \ha line. In studies with images at a high resolution, a morphological definition of the \hii regions using automatised tools is performed \citep{1994ApJ...431..156W, 2000A&A...363....9Z, 2002A&A...386..801Z, 2007ApJ...661..801O}. Nevertheless, these studies have been constrained only to less than five face-on and edge-on galaxies in their sample, based only in \ha in those studies with high resolution and with low resolution in those with spectroscopic information. However, the use of the IFS in recent years has allowed for the adoption of methodologies for DIG subtraction and analysis that are based on spatially-resolved spectroscopic information with broad samples, (e.g. MaNGA, CALIFA, or PHANGS). 

For example, DIG studies using data from the MaNGA survey (\citetalias{2017MNRAS.466.3217Z}, \citealt{2017A&A...599A.141J}) made use of 365 nearly face-on star-forming galaxies with 2.5 arcsec PSF FWHM, which is not enough to resolve individual \hii regions which typical sizes are $\sim$100 pc. In this particular case, \citetalias{2017MNRAS.466.3217Z} performed a cut out on \SBha$>10^{39}$ erg$\cdot$ s$^{-1}\cdot$ kpc$^{-2}$ to select those spaxels dominated by \hii regions emission. On the other hand, the studies based on CALIFA data \citepalias{2018MNRAS.474.3727L}, made use of 391 galaxies with a median PSF FWHM of $\sim0.8$ kpc and proposed an alternative method to separate the DIG and the \hii regions using the \ha equivalent width (\EWha), since (according to the authors) this parameter is a more appropriate proxy to distinguish the fundamental differences between the DIG and the star forming regions in comparison with the \SBha. In addition, the authors argue that the usage of \SBha could lead to misclassifications of low-surface-brightness \hii regions such as DIG.

Other authors have used the criteria to define and classify the DIG regime based on \EWha\, continued to be used to define the DIG \citep{2019MNRAS.489.4721V, 2020MNRAS.494.1622E}. However, redefining the DIG a priori as the ionised gas consistent with ionisation from HOLMES could be problematic if the main goal is to discern the different ionisation mechanisms \citepalias{2022A&A...659A..26B}. This is a challenging problem, as the standard resolution of CALIFA or MaNGA can not discern individual \hii regions in order to characterise the DIG morphologically. In addition, the high spatial resolution of some galaxies observed by IFS (e.g. M83-TYPHOON/PrISM; \citealt{2019MNRAS.487...79P}, M83-MUSE; \citealt{2022A&A...660A..77D}, PHANGS-MUSE; (\citetalias{2022A&A...659A..26B}, \citealt{2023A&A...672A.148C})), enables the use of automated tools in order to detect and remove individual \hii regions in IFU datacubes with resolutions of $\sim50$ pc, but constrained to a few objects.

In our work, we outline the approach for determining the DIG by utilising the spectroscopic and morphological data generated by the MUSE-IFS, using a broad dataset with high resolution and with spectroscopic information, not only \ha. A summary of the methodology, designed for galaxies of inclinations below $45\degree$, is provided below and it is subsequently explained in further detail in the following sub-sections:

    First,  to increase the S/N of the weak, low-surface brightness emission lines involved in the DIG study, a modified version of the adaptive binning technique from \citet{2023MNRAS.518..286L} is performed to the S/N map of the \sii line, obtaining then a segmentation map of the galaxy. The segmentation map is then applied to the datacube, to obtain a binned datacube, in which each bin corresponds to the integrated spectra as the sum of the spaxels contained in the bin.

    Then, a spectral fitting is performed on the binned spectra in order to derive emission line maps of the most important species for the DIG study. We considered nine species: the hydrogen \ha\, and \hb\, Balmer recombination lines, the \hei\, $\lambda5876$ recombination line, and the collisionally excited forbidden lines: \oiii $\lambda5007$, \oi $\lambda6300,$ and the doublets \nii $\lambda6548,6584$ and \sii $\lambda6717,6731$. The \ha and \hb equivalent widths (EWs) are also calculated in this step.

Finally, the DIG is separated from the emission of the \hii regions using a combination of an automated tool to detect and subtract the \hii regions from the binned \ha maps and a cut-off in the \ha surface brightness (\SBha) to subtract bright, irregular \hii regions not detected by the automated tools. This results in a mask that corresponds to the lower-limit of the DIG. This mask is applied to all the binned emission line maps, creating a set of the lower-limit DIG emission. The upper-limit DIG emission is derived from the lower-limit binned maps considering a constant DIG emission column above the projected areas of the HII regions. We will generalise the methodology for inclinations above $45\degree$ and edge-on galaxies in future papers. 

\subsection{Adaptive binning}

Typically, the distinction between SF regions and DIG has been made using \ha, but if we aim to explore the DIG in using all available spectroscopic information, we must consider key emission lines in the study of the DIG, such as \oi, \nii, \oiii, and \sii.
For a reliable analysis of the DIG, is crucial to take into account S/N limitations of the data, specially in those lines of lower surface brightness, such as those mentioned above.

The signal of an emission line feature of a spectrum can be defined as the difference between the maximum flux value of an emission line $f(\lambda_{em})$ centred on $\lambda_{em}$ of width $w_{em}$ (in \r{A}), and the mean of the fluxes in the two adjacent pseudo-continuum bands $f(\lambda_{c1})$ and $f(\lambda_{c1})$ of widths, $w_{c1}$, $w_{c2}$, measured on the spectrum of a given spaxel; and the noise corresponds to the mean of the flux within the two adjacent pseudo-continuum bands, $f(\lambda_{c1})$ and $f(\lambda_{c1})$ \citep{2012A&A...539A..73R}.


For the sake of clarity, in this work, the S/N of an emission line is the ratio of the amplitude of the emission line, defined as the peak of the line minus the mean flux of the pseudo-continuum, over the standard deviation of the adjacent pseudo-continuum on the spectrum: 

\begin{equation}
    S/N(\lambda_{em})=\frac{\mu}{\sigma}=\frac{f(\lambda_{em})-\langle{f(\lambda_{c1}),f(\lambda_{c2})}\rangle}{\sqrt{\sigma^2(f(\lambda_{c1}),f(\lambda_{c2}))}}
    \label{eq:SNratioDefinition}
.\end{equation}

The most commonly employed approach to enhance the S/N ratio of the data is to conduct adaptive binning, where individual pixels are combined into larger entities known as "bins" until the desired target S/N is achieved, 
However, this comes at the expense of diminished spatial resolution. The Voronoi binning method \citep{bookVoronoi,2003MNRAS.342..345C} generally resolves the issue of retaining the
highest spatial resolution of the images while adhering to the minimum S/N limitation.
This is achieved by tessellating the image and adjusting the bin
size to ensure that every bin attains the desired S/N. However, this technique is particularly well suited for analysing elliptical and featureless galaxies as it relies on the continuum S/N. It may not be as effective for our purpose, as our maps show irregular structures, such as those depicting nebular gas emission lines. Consequently, we require an alternative method to avoid losing the primary morphology of spiral galaxies.

\begin{figure*}[ht!]
    \hspace*{-5mm}
    \includegraphics[width=0.35\textwidth]{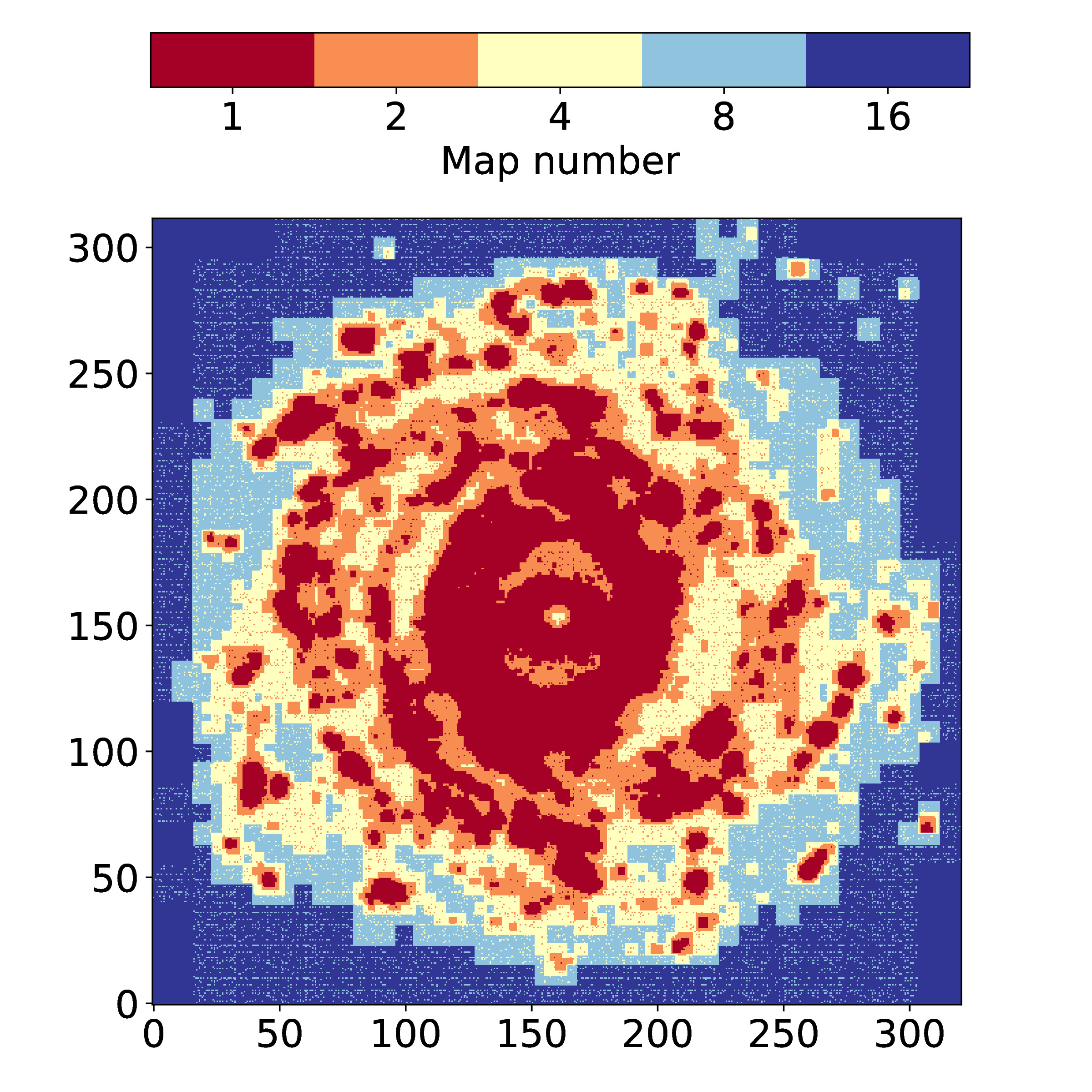}
    \hspace*{-5.5mm}
    \includegraphics[width=0.7\textwidth]{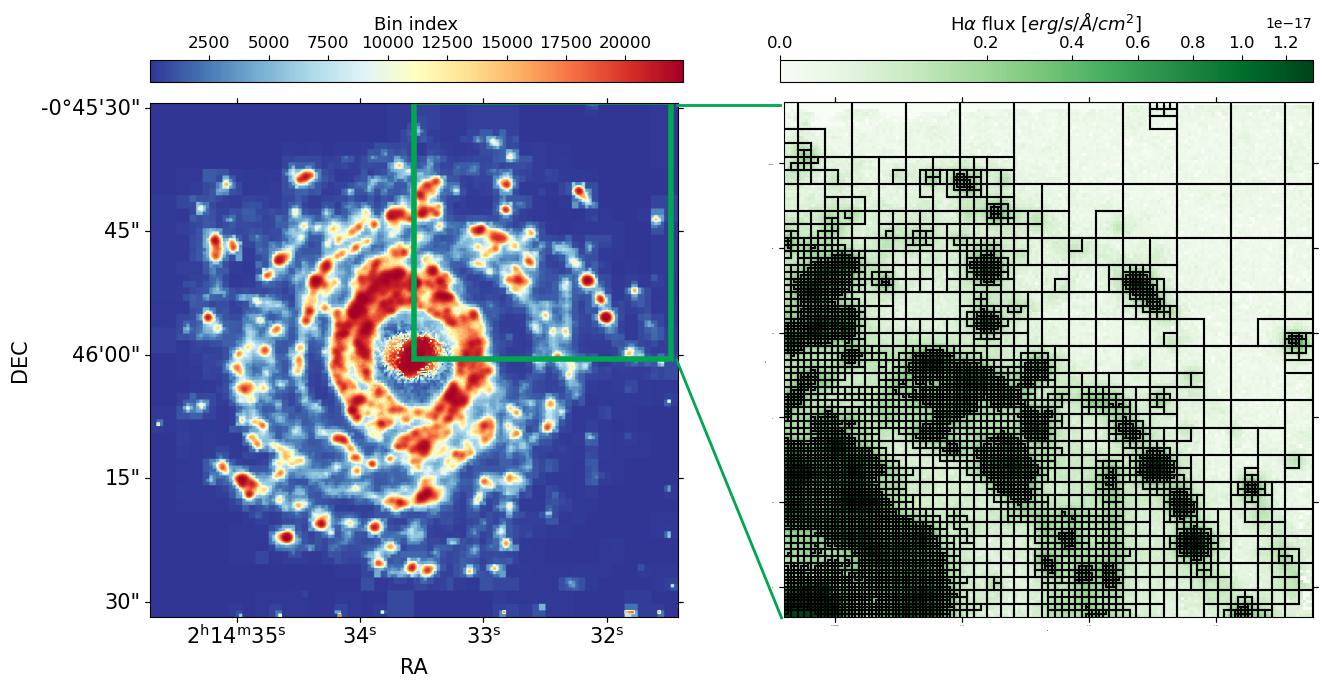}
    \caption{Example of an adaptive binning of NGC863 using $S/N(\sii)=10$ as target. The first figure represent the map numbers ($map_N$) from \citet{2023MNRAS.518..286L} algorithm. The second figure is the segmentation map obtained from our algorithm, where the bin index goes from $k=1$ to $N_{bins}$. The third figure is a close-up of the segmentation map plotting the borders of the bins over the \ha\, map. It is noticeable that the \hii regions, whose S/N is higher, are not binned, maintaining the structure pixel by pixel, and the bins are getting bigger as we move out of the \hii regions.}
    \label{fig:segmaps}
\end{figure*}

To solve this, we used a modified version of the adaptive binning method, introduced by \citet{2023MNRAS.518..286L} in order to enhance the S/N of weak emission lines. The code takes the flux of the line, the noise, and target S/N as its input to create a series of maps that cover the same area as the input data (map$_1$,...,map$_N$,...,map$_{N_{max}}$), as seen in Fig. \ref{fig:segmaps}. For each map$_N$ the set of N$\times$N pixels are averaged, and, as a consequence, the S/N will be increased. If the S/N of the bin is not the target S/N, the code takes the map$_{N+1}$ value instead. In our algorithm, we modified this code to get a segmentation map for binning the observed cube and getting a set of binned emission line maps (one per each considered line). The new binning technique was carried out as follows:

    First, we take as input the signal and noise defined in Eq. \ref{eq:SNratioDefinition} and a target S/N. The code then creates the series of map$_N$; however, instead of recovering the new averaged flux, we save an index, \textit{k,} in the coordinates [i$_0$, j$_0$], [i$_0$, j$_1$], [i$_1$, j$_0$], [i$_1$, j$_1$],...,[i$_N$, j$_N$], whose flux can then be averaged. The next subset of coordinates whose flux will be averaged will have an index of k+1.
Then, we perform the previous step for all subsets of coordinates for k = 1,...,N$_{bins}$. This  results in a segmentation map, with the lowest bin indices k corresponding to the higher bins, namely, the 1$\times$1 bins that correspond to the pixels with S/N higher than the S/N target and the lower indices corresponding to the bigger bins, namely, those needing more pixels to reach the target S/N (as seen in Fig. \ref{fig:segmaps}).
Finally, for each bin of the segmentation map, we save the integrated spectra of the observed cube as the sum of the spaxels contained in the bin. This result in a binned observed datacube.

Binning the observed cube offers the advantage of enhancing the S/N of the spectra, rather than merely reducing the relative error associated to the Gaussian fitting of the lines, as is the case with direct binning the emission line maps. Moreover, if our goal includes generating binned EW maps, it is not methodologically correct to bin the EW map obtained through spaxel-by-spaxel fitting. This is because EW is not an additive quantity; it varies when there are changes in the underlying stellar continuum. Hence, it is important to calculate the new EW from the integrated spectra.
If we want to increase the S/N of all the lines of interest, we need to choose a target feature as a basis to construct the segmentation map for the observed cube, ensuring to recover the maximum spatial information and signal in the DIG regimes. We employed S/N(\sii) as the target for adaptive binning because it is a low-ionisation collisional excitation line that remains unaffected by the correction for stellar population after the spectral fitting process.

\begin{figure*}[ht!]
    \includegraphics[width=0.3\textwidth]{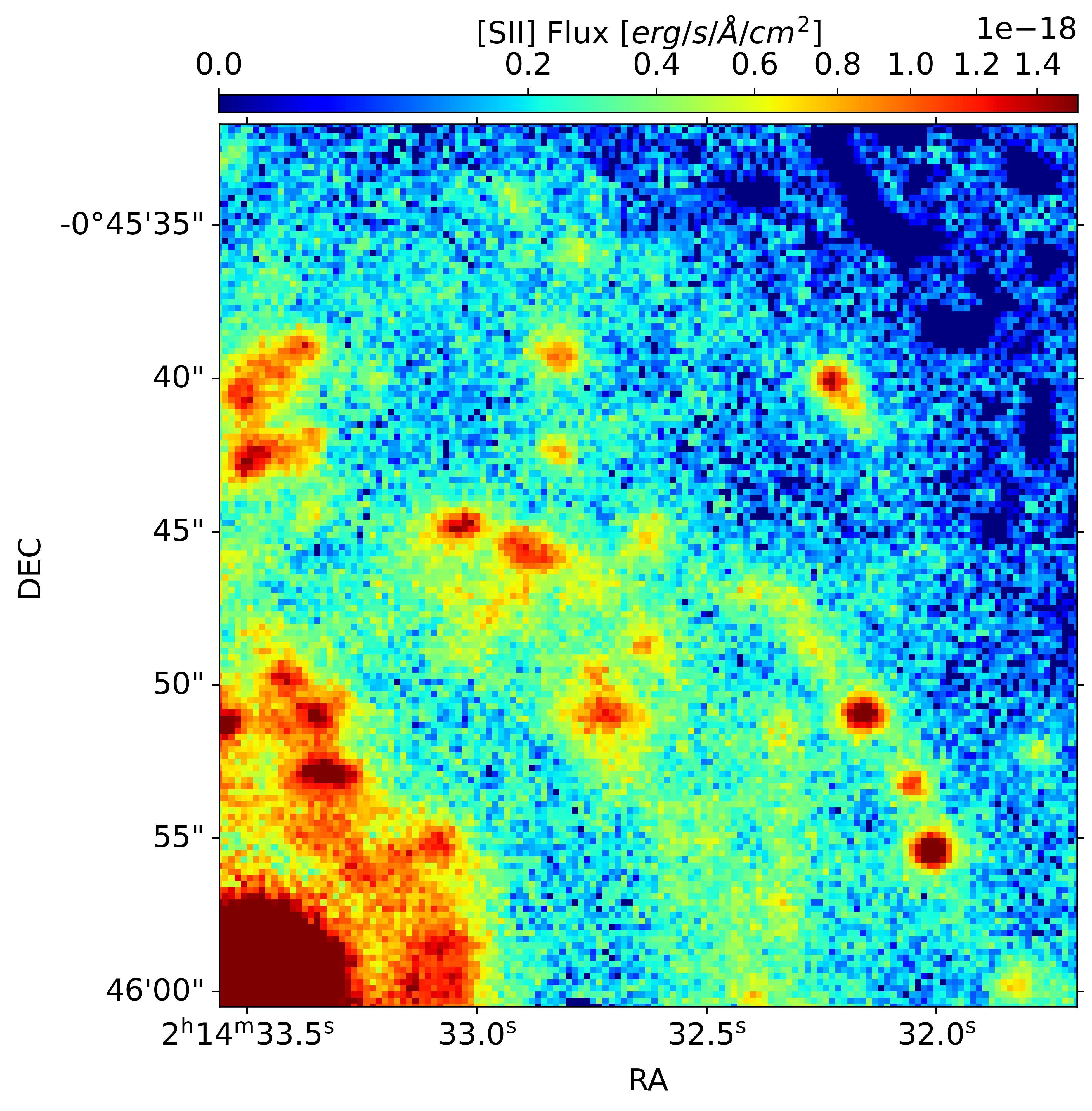}
    \includegraphics[width=0.3\textwidth]{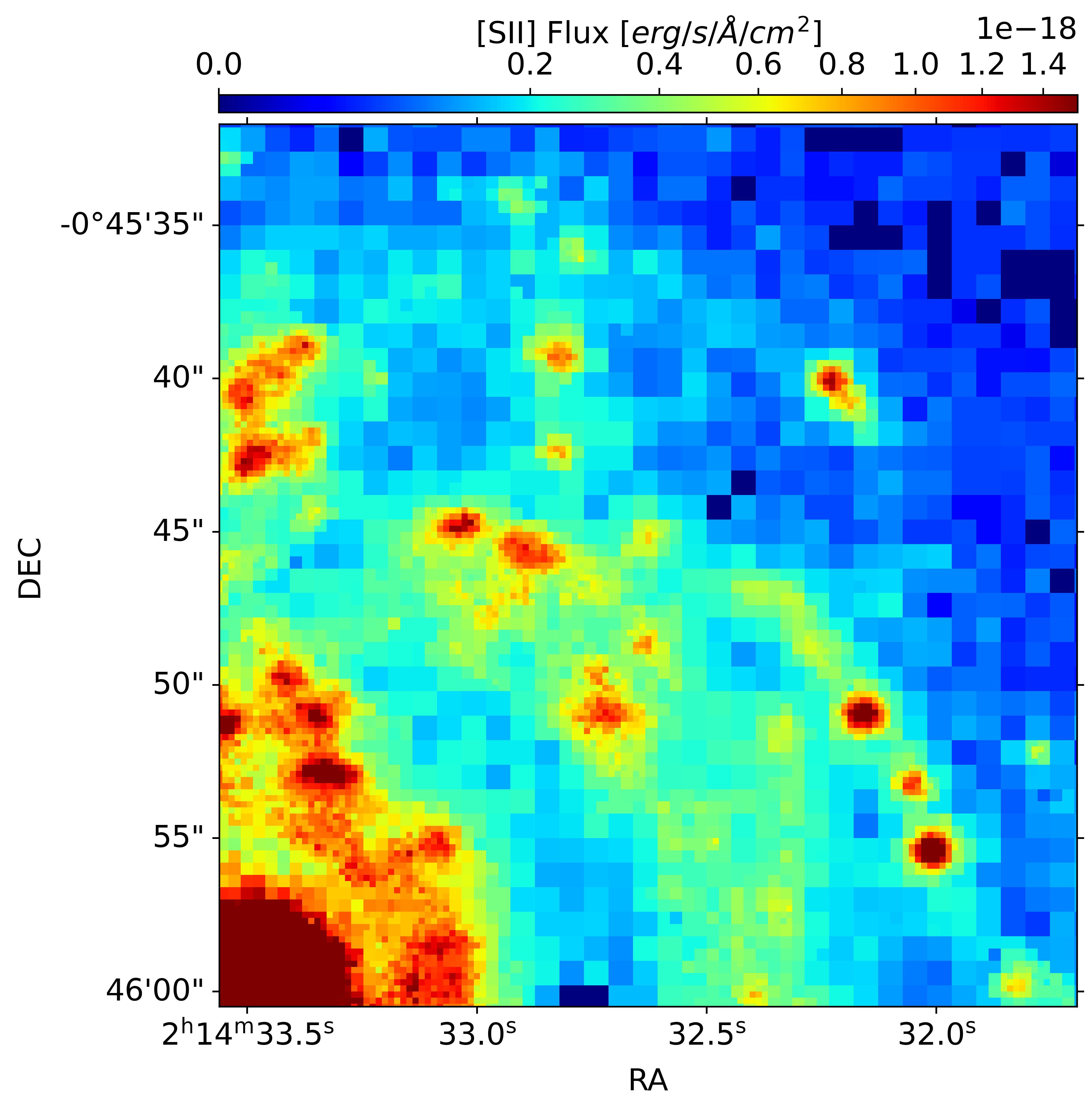}
    \includegraphics[width=0.4\textwidth]{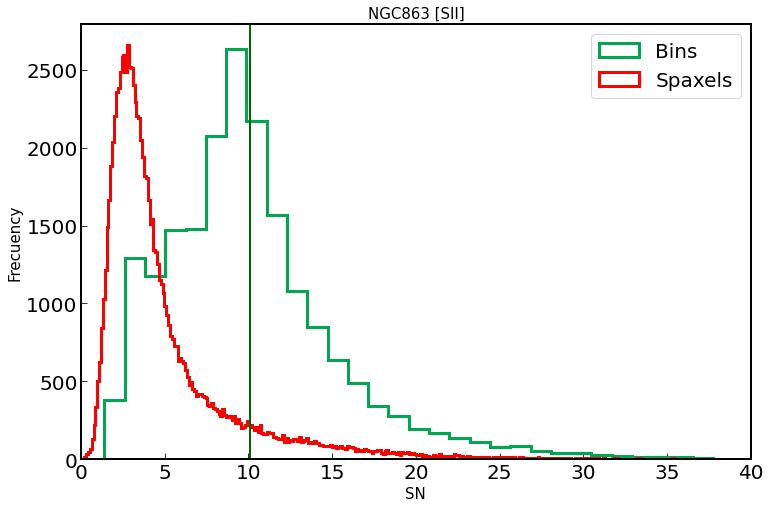}
    \caption{NGC863. Left panel: \sii emission line map obtained from the Gaussian fitting of the \sii line from the nebular gas cube spaxel-by-spaxel. The nebular cube was obtained as the difference between the observed cube and stellar cube as described in Appendix \ref{sec:SSP_apendA}.
    Central panel: Binned \sii emission line map, from the binned nebular gas cube, after performing an adaptive binning with target $S/N(\sii)=10$ to the observed cube, as described. Both panels are close-up of the same region than Fig. \ref{fig:segmaps}.
    Right panel: Distributions of S/N  of the \sii line for NGC863, calculated with the equation \ref{eq:SNratioDefinition}. Red distribution corresponds to the $S/N(\sii)$ measured in the observed cube, namely, the $S/N(\sii)$ of the spaxels. Green distribution corresponds to the $S/N(\sii)$ measured in the binned cube, namely, the $S/N(\sii)$ of the bins. The vertical dark green line marks the mean value of the S/N([\sii]) of the bins, showing that the adaptive binning technique fulfils the goal of reaching a target $S/N(\sii)=10,$ on average. }
    \label{fig:SIIexamples}
\end{figure*}

Considering the most important low-ionisation species that are found to be enhanced in the DIG regimes in SF galaxies, such as \nii and \sii \citepalias{2017MNRAS.466.3217Z, 2018MNRAS.474.3727L}, the \sii line exhibits the lowest S/N (mean of 5, median of 3) among the low-ionisation species. Furthermore, the \oiii line exhibits a mean S/N of 2.5, therefore, if a $\sim5\sigma$ detection is required, the S/N needs to be increased by a factor of 7 \citepalias{2022A&A...659A..26B}. To achieve this, we performed our adaptive binning method using S/N(\sii) = 10 as the target S/N. Figure \ref{fig:SIIexamples} (right panel) shows the distribution of S/N(\sii) obtained in the integrated spectra, where we can find that we get the average S/N of 10 that we expected. 

\begin{figure*}[h!]
    \includegraphics[width=\textwidth]{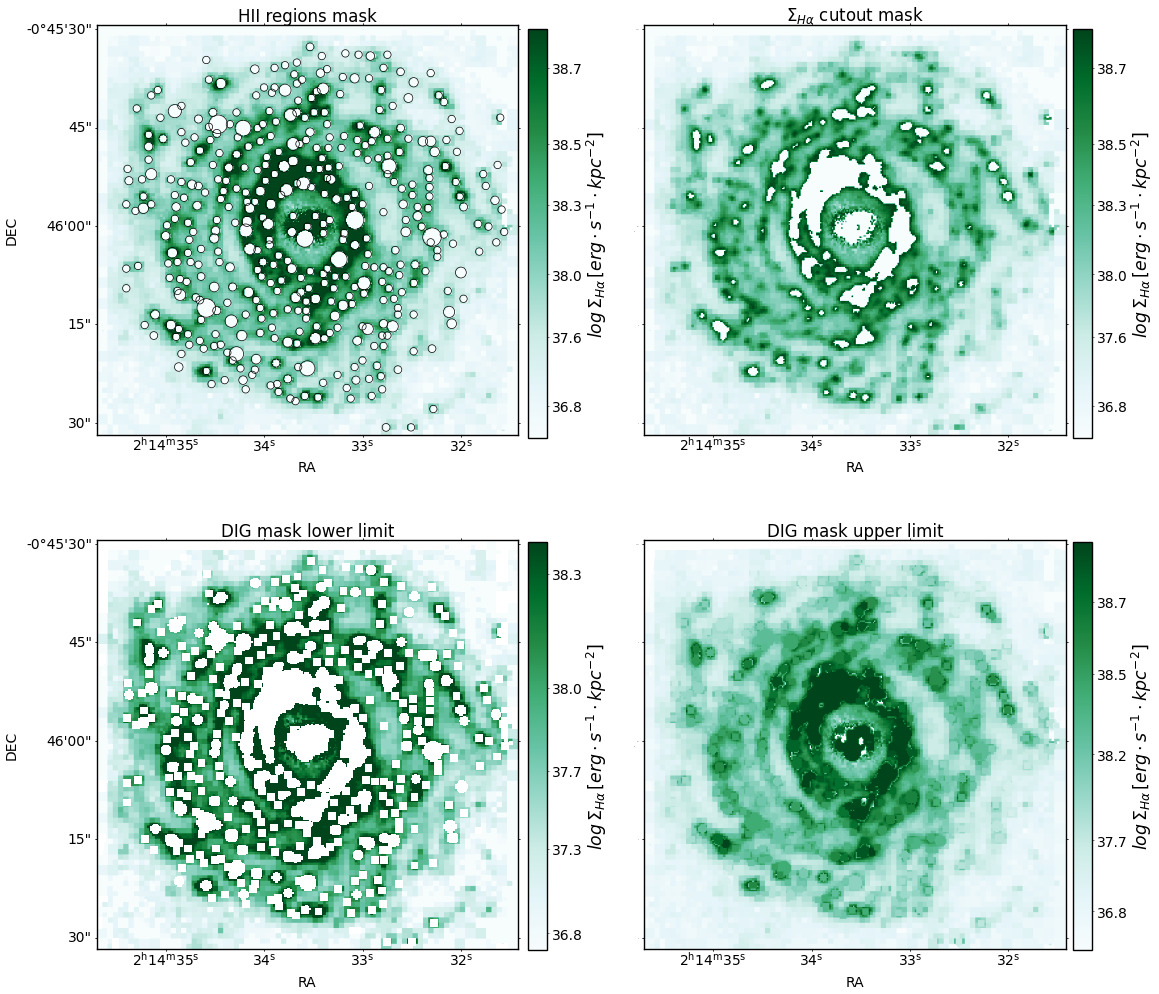}
    \caption{Steps to get the lower and upper DIG limit for NGC863 galaxy. The upper-left part corresponds to the DIG mask obtained by masking the \hii candidates given by the centroids and radii obtained from {\sc pyHIIextractor} to the binned \ha map. The upper-right part corresponds to the binned \ha map after applying a $3\sigma_{\Sigma(H\alpha)}$ cut-off. The lower-left part is the mask of the lower limit of the DIG, obtained as the combination of the upper-left and upper-right masks. The lower-right part corresponds to the upper limit, assuming a constant non-zero flux level above the \hii regions, interpolating the flux around the these regions.}
    \label{fig:DIG_limits}
\end{figure*}

\begin{figure*}[ht!]
    \includegraphics[width=\textwidth]{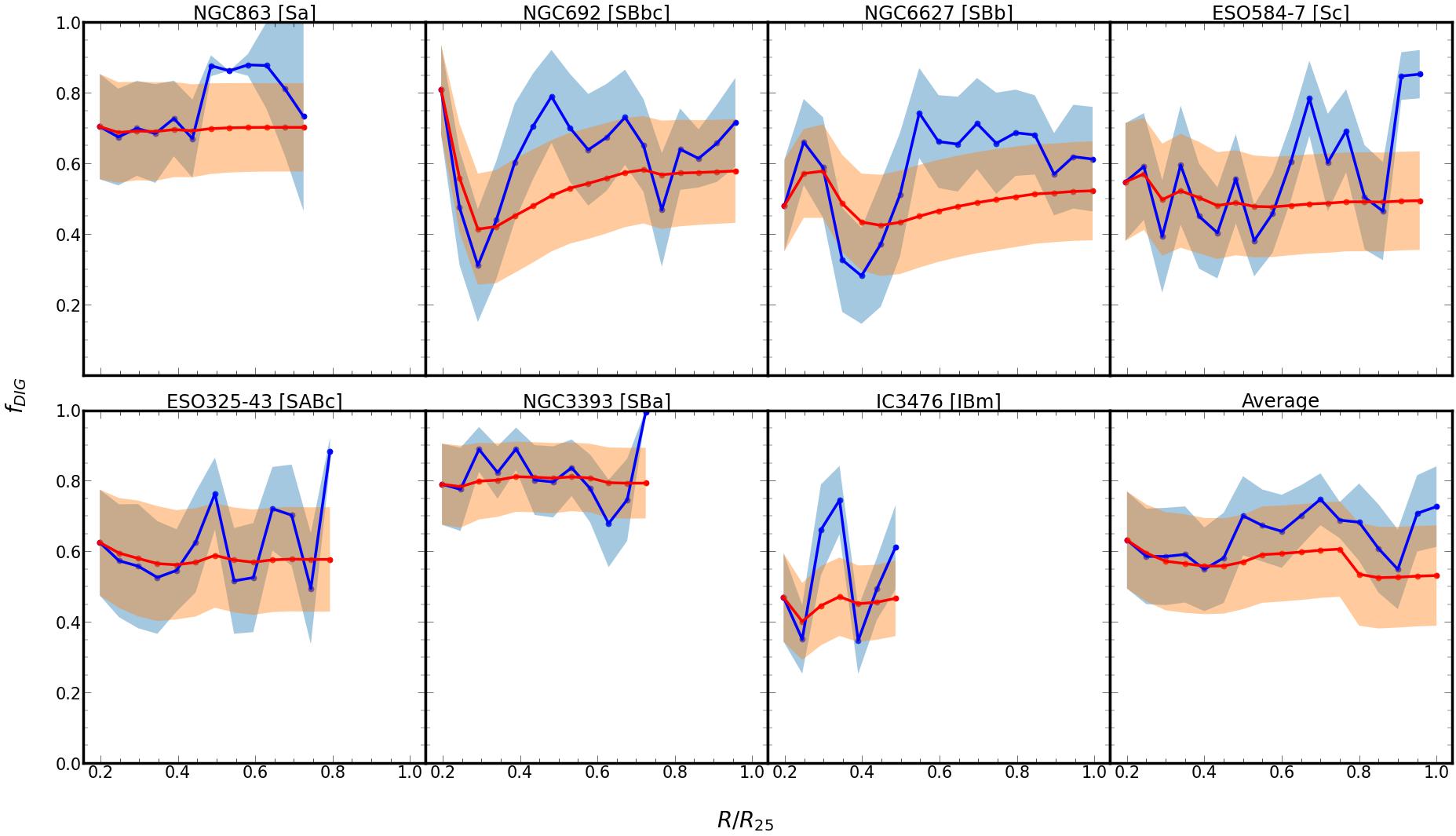}
    \caption{Distribution of the DIG fraction (f(\ha)$_{DIG}/$f(\ha)$_{total}$) for each galaxy of the sample. Red area represents the cumulative distribution between the the upper and lower DIG limits, with a step of 0.05R$_{25}$, being the solid red lined the mean value. Blue area is the radial distribution of the DIG obtained integrating the flux between rings of width $0.05R_{25}$ and with a step of 0.05R$_{25}$, being the solid blue lined the mean value. R$_{25}$ is the isophote at the blue brightness of 25 mag/arcsec$^2$. This parameter is obtained from Hyperleda database.}
    \label{fig:f_dig}
\end{figure*}

\subsection{Spectral fitting}\label{subSEC:fitting}
For every new spectrum, we performed a SSP fitting with {\sc STARLIGHT} \citep{2005MNRAS.358..363C}, with the obtained nebula emission spectra being the difference between the integrated observed spectra and the synthetic stellar spectra obtained after the fitting. We used the CB07 base spectra for the fittings \citep{2003MNRAS.344.1000B, 2007ASPC..374..303B}.
The number of SSPs (N$_*$ = 100) from CB07 comprises three metallicities, Z = 0.2 Z$_\odot$, Z$_\odot$ and 2.5 Z$_\odot$, and 15 ages from t = 0.001 to t = 13 Gyr. All SSPs are
normalised to 1M$_\odot$ at t = 0.
Their spectra were computed with Padova-2004 evolutionary tracks models, and \citet{2003PASP..115..763C} IMF (0.1M$_\odot<$ M $<$ 100 M$_{\odot}$).
Then we computed a Gaussian fitting around the nine lines of interest  on every nebular spectra, using the python library {\sc MPFIT}, resulting in a set of nine binned emission line maps. We refer to Appendix \ref{sec:SSP_apendA} for further details about the SSP fitting and emission line fits. Figure \ref{fig:SIIexamples} shows the difference between the \sii emission line map obtained after performing the SSP fitting on the observed cube spaxel-by-spaxel (left panel) and on the binned cube applying the exposed method (central panel).

The \ha and \hb EWs are also calculated during this step using the ratio between the integrated observed spectra and the fitted stellar model, resulting in a normalised stellar absorption spectra, where the EW is fitted for every bin using again {\sc MPFIT}.

\subsection{Morphological definition of the DIG limits}\label{subSEC:DIGdef}

The separation between SF regions and the DIG has been always the first topic of discussion when studying and analysing the physics of the DIG. Historically, the most common method has been to separate the DIG from SF regions based on a \ha\, surface brightness (\SBha) cut-off (\citealt{2000A&A...363....9Z, 2007ApJ...661..801O}; \citetalias{2017MNRAS.466.3217Z}). However, this method presents the problem of misclassifying low-surface-brightness \hii regions as DIG, in addition to the possibility of classifying the emission of two overlapped DIG regions as \hii regions \citepalias{2018MNRAS.474.3727L}. 

A suitable alternative for this problem is using automatised tools for the detection and subtraction of the \hii regions individually. Several tools have been developed for this task, including {\sc SExtractor} \citep{1996A&AS..117..393B}, {\sc HIIphot} \citep{2000AJ....120.3070T}, {\sc HIIexplorer} \citep{2012A&A...538A...8S}, {\sc pyHIIexplorer} \citep{2020MNRAS.494.1622E}, {\sc astrodendro} \citep{2022A&A...660A..77D}, {\sc pyHIIextractor} \citep{2022RASTI...1....3L}, and {\sc CLUMPFIND} \citep{2023A&A...672A.148C}.

The spatial resolution is an important key to detect and define morphologically individual \hii regions. For the specific range of resolutions of the BETIS sample, we used {\sc pyHIIextractor} \citep{2022RASTI...1....3L} in order to detect \hii regions candidates. The code detects the \hii regions candidates and assigns them a radius and a centroid, in the image coordinates. It is highly efficient for the detection of circular \hii regions, especially those with low \SBha, solving the aforementioned problem. Nevertheless, the limiting factor of this algorithms is in the complexity of the regions and in the variety of shapes and sizes of the brighter ones; hence, a circular extraction is not sufficient for those complex regions, since it would not be adapted to their morphology.

\begin{figure*}[ht!]
    \includegraphics[width=\textwidth]{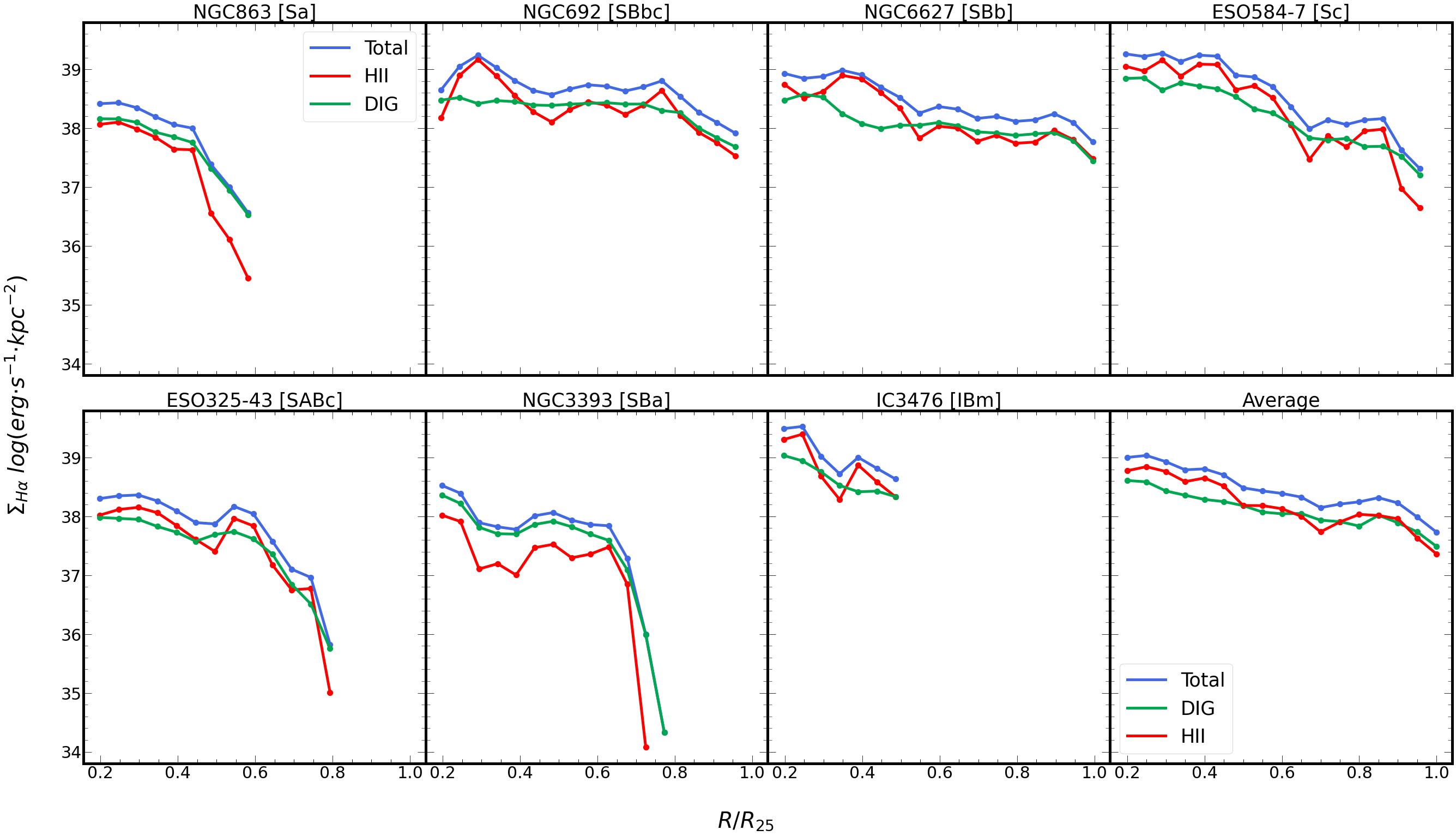}
    \caption{Radial distribution of \SBha for each galaxy, obtained by measuring the surface brightness of a series of de-projected annulus centred at the nucleus of each galaxy with a width of $0.05R_{25}$. The distributions are performed for the entire galaxy (blue), the lower DIG limit (green) and HII regions (red). The last panel represent the mean distributions for the sample.}
    \label{fig:SB_radial}
\end{figure*}

In the pursuit of the optimal methodology for segregating DIG emissions from \hii regions, we conducted an extensive series of tests employing a variety of galaxies, techniques, and algorithms, considering the different observing conditions and depth of the BETIS sample. Our findings have led us to the conclusion that to establish a reliable morphological definition of the DIG, a combination of two methods is necessary:
(i) the detection and masking of bona-fide \hii regions exhibiting regular (circular) morphology based on the \ha emission line map; and (ii) the implementation of a cut-off in \ha surface brightness to account for irregular and highly luminous \hii regions that may elude automated masking tools.

For the first step, we made use of the code {\sc pyHIIextractor} for a first detection of the \hii regions. The algorithm takes a \ha map and detects the positions and radii of all the candidates \hii regions. We used those positions and radii to perform a mask of \hii regions,  applied to the binned \ha map. 

For the second step, we performed a \SBha cut-off to the binned map. This cut-off is defined as three times the standard deviation of the surface brightness distribution of the masked map, as defined in the initial step ($3\sigma_{\Sigma(H\alpha)}$). This cut-off corresponds to three times the average \ha background level of the showcase sample.
Performing this cut-off will result in a second DIG mask, that in combination with the first step DIG mask will give us the lower limit of DIG, as seen in Fig. \ref{fig:DIG_limits}, since we are assuming that all the emission coming from the \hii positions is due to the star formation and the DIG contribution can be neglected.

We can also estimate an upper limit for the DIG, rejecting the previous assumption and considering a constant DIG emission column above the projected areas of the \hii regions \citep{2000A&A...363....9Z, 2023A&A...672A.148C}. This limit is estimated in two steps. Firstly, for every \hii region of centroid ('X','Y') and radius 'R' detected by {\sc pyHIIextractor}, we defined an annulus centred in that detection, with inner radius equal to the radius obtained and outer radius 1.4 times the inner radius. We filled the \hii region with the mean surface brightness measured inside the annulus. After performing this 'filling' with every \hii region detection, part of the galaxy is still masked due to the \SBha cut-off. To estimate the DIG contribution in these remaining regions, we filled them with the mean value of surface brightness of the border of the region. This result in a upper limit \ha DIG map, as seen in Fig. \ref{fig:DIG_limits}. The lower-limit DIG mask is then applied to the rest of the binned emission line and EW maps, which is the basis of our data. 

\section{BETIS: first results}\label{SEC:results}

\begin{table}[!t]
\centering
\resizebox{\columnwidth}{!}{%
\begin{tabular}{lllllll}
\hline
\hline
Line & \ha & \hb & \nii & \sii & \oi & \oiii \\ \hline
\% S/N (spax) & 65 & 27 & 50 & 41 & 11 & 51 \\
\% S/N (bin) & 96 & 90 & 93 & 91 & 63 & 82 \\ \hline
\end{tabular}%
}
\caption{First column is the percentage of spaxels and bins of the sample whose S/N(\ha) is higher than 3. The rest of the columns represent the percentage of the spaxels and bins with the S/N of \ha and where the S/N of the corresponding line higher than 3.}
\label{tab:SNs}
\end{table}

After performing the binning methodology outlined above, we found an average bin size of the sample of 627 pc (see Table \ref{tab:SNs} for a overview of the increment of the S/Ns after binning). The average bin size in our sample is larger than the typical bin sizes in studies of narrowband \ha images (e.g. $\sim250$ pc, \citealp{1996AJ....111.2265F}; $\sim35$ pc (native), \citealt{2000A&A...363....9Z}; $\sim475$ pc, \citealt{2007ApJ...661..801O}). Moreover, the bin dimensions exceed those of the mean Voronoi bins of the PHANGS-MUSE sample ($\sim 200$ pc; \citetalias{2022A&A...659A..26B}). Nonetheless, they are smaller than the average PSF FWHM of the CALIFA galaxies ($\sim 800$ pc; \citetalias{2018MNRAS.474.3727L}).

To ensure the robustness of our analysis, we exclusively considered DIG bins with a relative error of 40\% or lower and a S/N greater than 3 for all the lines involved in the subsequent analyses of this section. We also exclude the central part of those exhibiting an AGN: NGC863, NGC692 and NGC3393. Those constraints leave us with 80\% of the DIG bins.

\subsection{The DIG fraction}

The visual inspection and variation of the DIG, both within individual galaxies and across them, provide valuable insights into its origins. In the binned \ha maps of our showcase sample galaxies, a diverse range of structures within the ionised gas regions becomes evident. The variations in \ha emissions in different directions yield crucial information about the evenness on the intensity of the DIG distribution and its spatial association with prominent \hii regions. Likewise, investigating changes in the diffuse fraction and \ha surface distribution within a particular galaxy is of significant interest as it aids in pinpointing the source of the DIG.

Figure \ref{fig:f_dig} shows the distribution of the DIG fraction, defined as the ratio of the DIG flux to the total \ha flux,  f(\ha)$_{DIG}$/f(\ha)$_{total}$, for each galaxy of the sample, in both the radial and cumulative distributions.
The radial distribution is obtained by measuring the DIG fraction from a series of de-projected rings centred at the nucleus of each galaxy with a width of 0.05R$_{25}$, while the cumulative distribution is obtained by performing aperture photometry of the DIG fraction in the same series of de-projected rings that are used for deriving the radial profiles, for both the lower and upper DIG limits. Moreover, using the definition of radial distribution, we present in Fig. \ref{fig:SB_radial} the \ha surface-brightness  distributions for the lower DIG limit (green) after applying the mask (defined in Sect. \ref{subSEC:DIGdef}) for the galaxy with only \hii regions (red) and the total galaxy (blue). All the fluxes from this section and hereafter have been corrected for interstellar extinction assuming the Cardelli extinction law, assuming R$_V$ = 3.1 \citep{1989ApJ...345..245C} and a Balmer decrement \ha/\hb = 2.85.

Overall, the radial distribution of the DIG fraction tends to increase, while the cumulative distributions remain constant, but it is more affected by the galaxy morphology than the cumulative distribution. The presence of a barred structure in NGC692, NGC6627, and ESO325-43 is evident through a decline in both distributions between 0.2-0.4 R$_{25}$, which decreases up to 20\%. This effect is also reflected on the radial distributions of the \ha surface brightness (Fig. \ref{fig:SB_radial}), which show an increment of the \SBha for the \hii regions distribution, reaching up to 10$^{39}$ erg$\cdot$ s$^{-1}\cdot$ kpc$^{-2}$ between 0.2-0.4 R$_{25}$ for those galaxies.
NGC3393 exhibits the highest DIG fraction for both lower and upper limits (0.69-0.87), presenting a surface brightness that surpasses the \hii regions, throughout the entire galaxy, resulting in a maximum difference of around 1 dex. Furthermore, NGC863 also exhibits higher \SBha in the DIG than in the \hii regions, notably towards greater galactocentric radii. 

The Sc galaxy ESO584-7 and the dwarf IBm galaxy IC3476 present the lower DIG fraction, 0.37-0.66 and 0.4-0.63, respectively. Additionally, they have the highest surface brightness for the \hii regions in the sample, being >10$^{39}$ erg$\cdot$ s$^{-1}\cdot$ kpc$^{-2}$ in the inner parts of the galaxy and decreasing to $\sim 5\cdot10^{38}$ at 0.5R$_{25}$. The average radial distribution of the DIG fraction is not showing any tendency in particular, remaining constant probably due to the mixture of morphologies of our showcase sample. A full analysis considering a general distribution by morphological type will be carried out as part of BETIS in a forthcoming paper.

Table \ref{tab:DIG_frac} summarises the results obtained from the DIG fraction for the showcase sample. The general findings for this showcase sample show that the DIG fraction ranges from 0.4 to 0.7, which coincides with the results of previous research, in terms of both the lower limit (e.g. \citealt{1996AJ....111.2265F, 1996AJ....112.2567F, 2000A&A...363....9Z, 2002AJ....124.3118T, 2007ApJ...661..801O}; \citetalias{2022A&A...659A..26B}), and upper limit (e.g. \citealt{2000A&A...363....9Z, 2023A&A...672A.148C}). In addition, the average \SBha cut-out in also consistent with the average upper DIG level of previous studies (\citealt{2019MNRAS.487...79P}, \citetalias{2022A&A...659A..26B}, \citealt{2023A&A...672A.148C}). In addition, the tendency of the cumulative distribution is to be constant and indicating approximately a 60\% of DIG in these galaxies. The surface brightness of the DIG vary from $\sim 6\cdot10^{38}$ in the inner parts of the galaxies and decreasing monotonically to 
$\sim 5\cdot$10$^{37}$ erg$\cdot$ s$^{-1}\cdot$ kpc$^{-2}$
in the outer regions, with a notably high integrated \ha luminosity between $10^{40}$ and $5\cdot10^{41}$ erg/s. This is consistent with previous studies (e.g. \citealt{1996AJ....111.2265F, 1996AJ....112.2567F, 2000A&A...363....9Z}).

The similarity in the \SBha radial distributions between the \hii and DIG regimes and its impact in the DIG fraction supports the correlation between \hii regions emission and DIG photoionisation \citep{1996AJ....111.2265F, 1996AJ....112.2567F, 2000A&A...363....9Z}. However, performing this exploration individually shows that there are instances where DIG \SBha values exceed those of the \hii regions, in particular, in the two Seyfert galaxies, NGC863 and NGC3393. For instance, for NGC863, a total DIG \SBha of $6.6\cdot10^{38}$ erg$\cdot$s$^{-1}\cdot$ kpc$^{-2}$ will require a power per unit area of 1.5$\cdot10^{-3}$erg$\cdot$s$^{-1}\cdot$cm$^{-2}$ to keep the DIG ionised, while the total \SBha of the \hii regions of 5.02$\cdot10^{38}$erg$\cdot$s$^{-1}\cdot$kpc$^{-2}$ provides a power of 1.1$\cdot10^{-3}$erg$\cdot$s$^{-1}\cdot$cm$^{-2}$, which is insufficient to ionise the entire DIG\footnote{We are assuming that the average number of photons per recombination is 0.46 for this estimation.}. 

Hence, it is imperative to contemplate alternative ionisation sources that can provide additional energy supply to the ISM apart from Lyman continuum photons escaping from \hii regions to keep the DIG ionised for those particular cases. The incorporation of collisional, low-excitation lines such as \nii, \sii, or \oi, as well as high-excitation lines such as \oiii and \hei in the analysis is essential for comprehensively investigating the diverse array of ionisation mechanisms that exist within the (ISM).

\begin{table}[ht!]
\centering
\resizebox{\columnwidth}{!}{%

\begin{tabular}{llllll}
\hline
\hline
\textbf{Galaxy} & \textbf{Type} & \textbf{$f_{DIG, low}$} & \textbf{$f_{DIG, up}$} & \textbf{$3\sigma_{\Sigma(H\alpha)}$} & \textbf{Bin size (pc)} \\  \hline
NGC863 & SA(s)a & 0.45 & 0.75 & 38.8 & 743 \\
NGC692 & SBbc & 0.44 & 0.73 & 39.1 & 578 \\
NGC6627 & SBb & 0.39 & 0.67 & 39.3 & 490 \\
ESO584-7 & Sc & 0.37 & 0.66 & 39.4 & 920 \\
ESO325-43 & SABc & 0.44 & 0.75 & 38.5 & 1129 \\
NGC3393 & SBa & 0.69 & 0.87 & 40.0 & 427 \\
IC3476 & IBm & 0.40 &  0.63 & 39.8 & 101 \\  \hline
Subsample & - & 0.40 & 0.70 & 39.3 & 627 \\ \hline
\end{tabular}%
}
\caption{Total DIG fraction for each galaxy. Here, '$f_{DIG, low}$' and '$f_{DIG, up}$' are the upper and lower limit of the DIG. $3\sigma_{\Sigma(H\alpha)}$ is the \SBha cut-off performed for each galaxy in units of log $erg\cdot s^{-1}\cdot kpc^{-2}$. 'Bin size (pc)' is the average bin size of the galaxy. The last row indicates the average values of the sample.}
\label{tab:DIG_frac}
\end{table}

\begin{figure*}[ht!]
    \includegraphics[width=\textwidth]{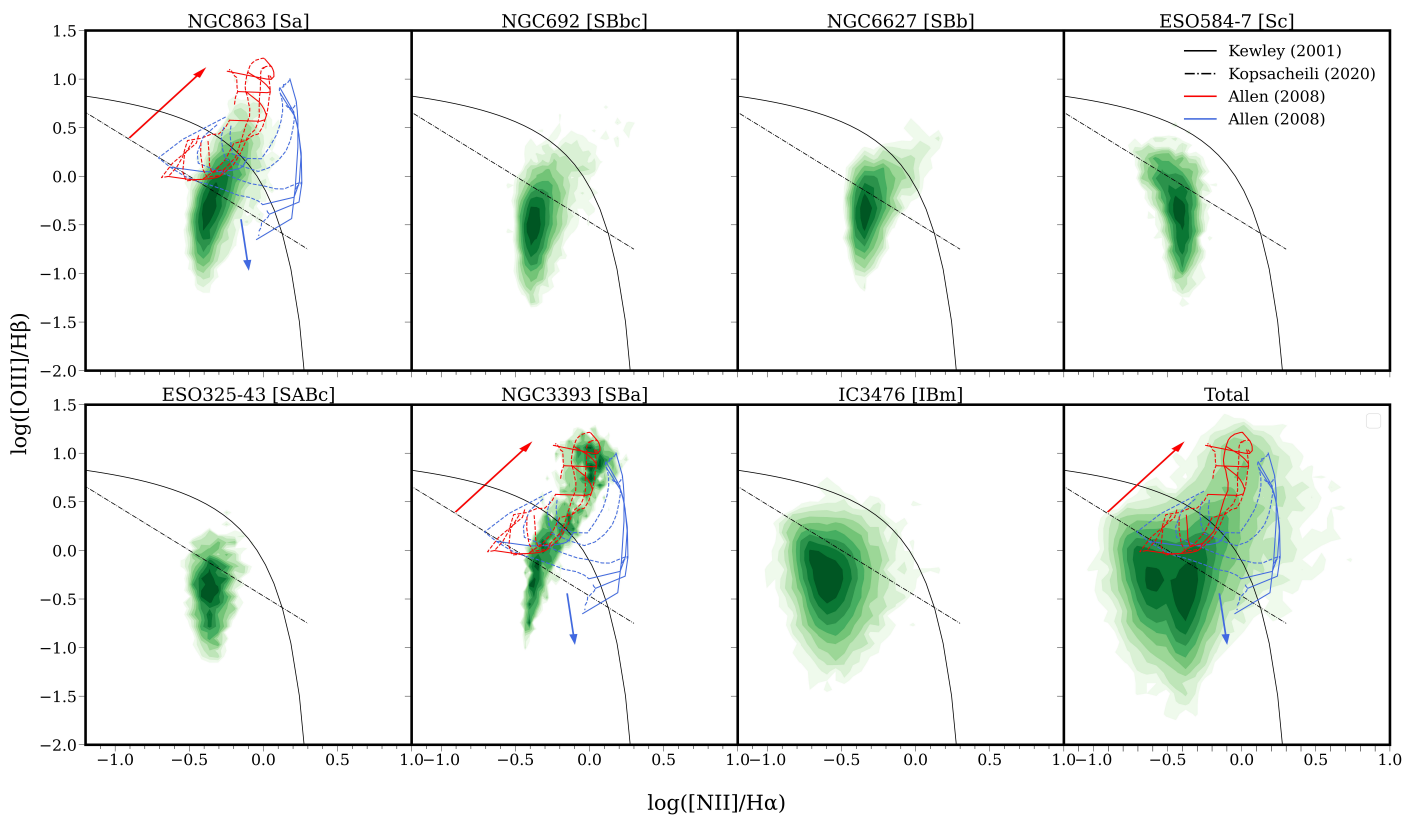}
    \caption{BPT Diagnosis of the sample for the DIG bins that verifies that the relative errors of \oiii line bins are below 40\% and its S/N above 3. The last panel displays the seven prior panels together. The central parts of NGC863, NGC3393, and NGC692 were excluded in all panels due to their strong AGN emissions. Each contour encloses the 10\% of the points, with every point a single bin. Black lines are given by \citet{2001ApJ...556..121K} (solid) and \citet{2003MNRAS.346.1055K} (dashed) for the classic \hii regions photoionisation and AGN demarcations. The dashed-dot line depicts one of the 2D diagnostics developed by \citet{2020MNRAS.491..889K} for the separation of shock excited (e.g. supernova remnants) from photoionised regions (e.g. \hii regions). Coloured lines represent the models of photoionisation
    by fast shocks from \citet{2008ApJS..178...20A}. The blue and red models illustrate photoionisation where only front shocks occur and when pre-ionisation by a precursor is taken into account. The solid model curves plotted represents shocks winds of 200, 400, 500 and 1000 km/s, and dashed model curves represents magnetic field intensities of 0.0001, 1.0, 5.0 and 10. Red and blue arrows represent the direction of increasing wind velocity in each model.}
    \label{fig:BPT_DIG}
\end{figure*}

\subsection{BPT diagnosis of the DIG}\label{SubSec:BPT_DIG}

We can explore the location of the DIG bins in the classical \nii BPT diagram, as shown in Fig. \ref{fig:BPT_DIG}. In the same figure we have included a diagnostic from \citet{2020MNRAS.491..889K} for the separation of shock excited from photoionised regions. The BPT of the global sample shows the DIG falls mostly below the \citealt{2001ApJ...556..121K} demarcation, showing a photoionisation feature due to \hii regions, but with high-excitation regions above the demarcation corresponding to AGN-like emission, as noted in previous studies and usually explained as photoionisation due to HOLMES (\citetalias{2018MNRAS.474.3727L, 2022A&A...659A..26B}). However, when performing this diagnosis for individual galaxies, only NGC863 and NGC3393, both Seyfert-2 AGNs, exhibit line ratios characteristic of AGN emission. All the DIG for the rest of the galaxies show photoionisation feature due to \hii regions.
The central region of NGC863 and NGC3393 is not considered in this analysis, so the AGN outflows could be the source of the gas ionisation with high-excitation lines found in the BPTs. We compare our results with the theoretical models of gas that is highly excited via fast-shocks of \citet{2008ApJS..178...20A}, assuming solar metallicity and a pre-shock density of 1 cm$^{-3}$. We plotted the predicted line ratios including (red lines) or not (blue lines) a photoionisation by precursor, and given a range of shock velocities (v$_s$) and magnetic field intensities (B). The line ratios observed in NGC863 are consisted with fast-shock without precursor with v$_s$ between 200 and 500 km/s and B between 0.0001 and 5. In the case of NGC3393, the ratios observed corresponds to fast-shock without a precursor, with v$_s$ between 200 and 1000 km/s and B between 0.0001 and 10. Furthermore, the fact that the ionisation bi-cone of NGC3393 and its continuum emission are uncoupled \citep{2016ApJ...829...46M}, along with the emission line ratios in the BPT diagram indicative of ionisation by fast shocks, suggests that we are tracing gas outflows rather than DIG
emission \citep{2020AJ....159..167L}.

The presence of AGN outflows affects the overall BPT diagnosis, revealing that a section of the DIG is ionised by sources separate from the photon leakage originating from \hii regions, but what we are introducing is the ionisation cone of the AGNs, mimicking the DIG emission. Therefore, when conducting a comprehensive diagnostic assessment of the DIG across a global sample, it is crucial to completely exclude galaxies that exhibit AGN emission. This occurs even if the whole DIG is ionised by star-forming regions, as observed in the remaining galaxies in the sample. This effect also explains the high \SBha in the DIG found in the same two galaxies; NGC863 and NGC3393, as we are incorporating the AGN emissions. 

\begin{figure*}[ht!]
    \includegraphics[width=\textwidth]{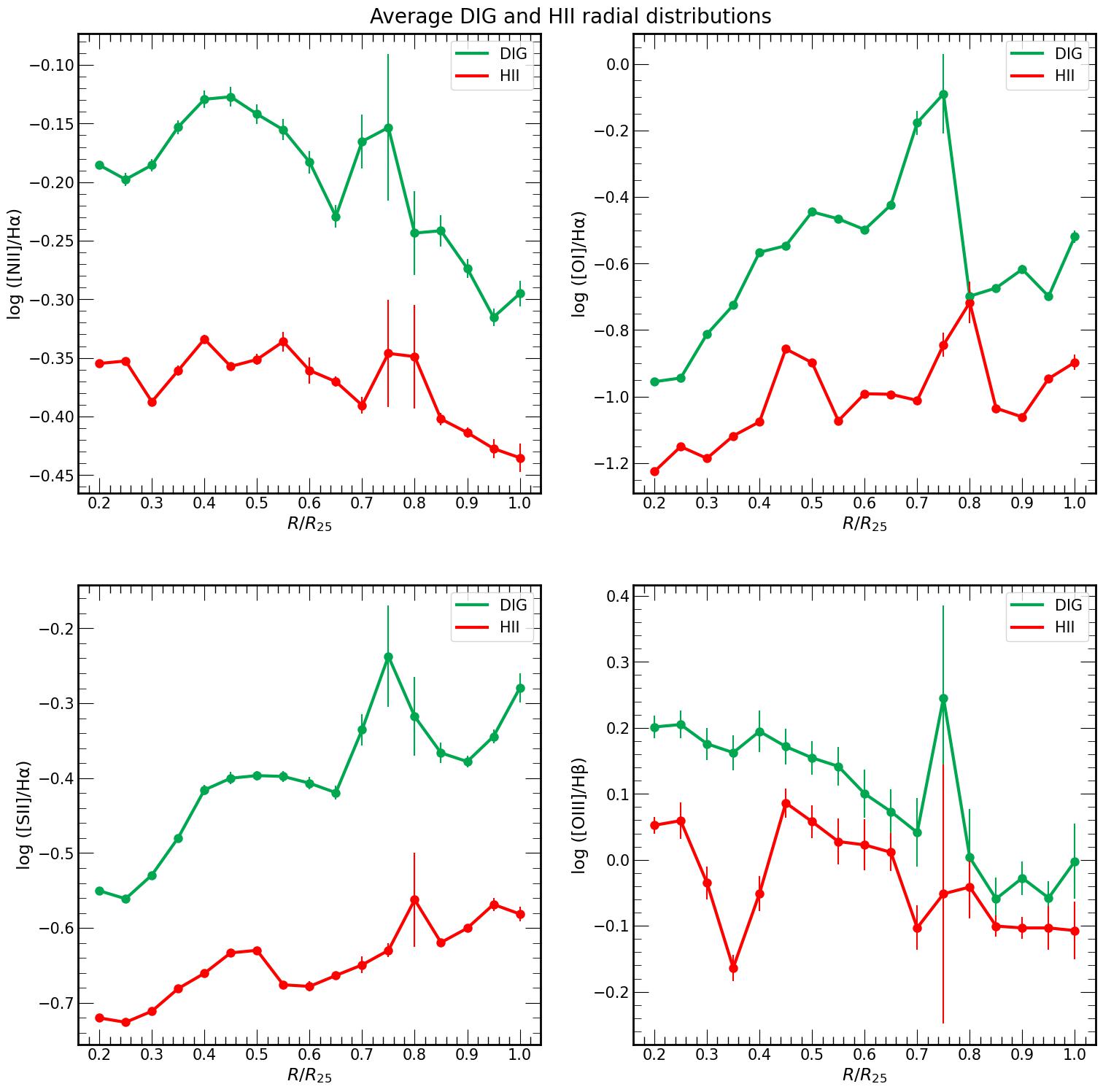}
    \caption{Radial distribution of the \sii/\ha ratio of the DIG bins (green), and \hii regions according to our definition (red), obtained as the ratio of the mean values of the respective fluxes within a ring of width $0.05R_{25}$ with a step of $0.05R_{25}$ from $0.2R_{25}$ to $R_{25}$. We are only considering those bins whose relative error in \sii and \ha\, lines are lower than 40\%. To avoid the strong AGN emissions, we start all the distributions from $0.2R_{25}$. In general we see a tendency to increase the ratio as we move to outer regions of the galaxy. Besides, the ratio is in all cases higher in DIG regions, as we expected from the literature \citep{2009RvMP...81..969H}.}
    \label{fig:mean_lines}
\end{figure*}

\subsection{DIG line ratios}\label{SubSec:line_ratios}

Further evidence of the connection between the DIG and \hii regions lies in the behaviour of the \sii/\ha, \nii/\ha, \oi/\ha , and \oiii/\hb ratios. Historically, studying the radial dependence of the DIG has been important, as its trend along different galactocentric distances can constrain its origin \citep{1996AJ....111.2265F}. Furthermore, recent investigations have highlighted the importance of considering possible contamination and/or contribution of the DIG to physical parameters derived from emission lines, such as metallicity and star-formation (\citetalias{2022A&A...659A..26B}, \citealt{2024MNRAS.tmp..356L}). Therefore, aperture (or, correspondingly, redshift) biases should be taken into account, for which the study of the DIG in a radial basis is a key element. An observational fact is that the low-ionisation \nii/\ha and \sii/\ha line ratios increase with decreasing \SBha, the best example being the increase in these line ratios with increasing distance from the mid-plane, both in the Milky Way \citep{1999ApJ...523..223H} and other galaxies \citep{1999ApJ...521..492R, 2000A&A...364L..36T}. Although the values for \nii/\ha and \sii/\ha could vary considerably in the DIG, they are correlated, often with a nearly constant \sii/\nii ratio over large regions \citep{2009RvMP...81..969H}.

Figure \ref{fig:mean_lines}, shows the average radial distribution of the \sii/\ha, \nii/\ha, \oiii/\hb, and \oi/\ha line ratios for the lower DIG limit. The distributions have been generated using the same method as for \SBha (as shown in Fig. \ref{fig:SB_radial}) and f$_{DIG}$ (blue curves of the Fig. \ref{fig:f_dig}) radial distributions. Each data point is computed as the average value from the seven galaxies in the sample at a specific radius. We created these distributions for both the DIG (in green) and the area corresponding to \hii regions (red). In all instances, we exclusively consider the bins with a relative error below 40\% for each line. In all cases, the DIG line ratios show higher values than the line ratios corresponding to the \hii regions.

The \nii/\ha radial distributions (upper-left panel), for both the DIG and the \hii regions, decrease with galactocentric distance. The \nii/\ha DIG distribution ranges from 0.67 to 0.51, while for the \hii regions from 0.44 to 0.37 between 0.2R$_{25}$ and 1R$_{25}$, being the \nii/\ha line ratio of the DIG 0.15 dex higher in average. The \nii/\ha line ratio is a well-know metallicity indicator \citep{1994ApJ...429..572S, 2002MNRAS.330...69D, 2015MNRAS.448.2030H}; thus, the decreasing value as a function of radius is primarily reflecting a change in metallicity of the ionised gas, in this case of the metallicity gradient of the spiral galaxies. Nevertheless, given that N$^+$/N and H$^+$/H vary little within the DIG, for a given metallicity, variations in the \nii/\ha line ratio essentially trace variations in $T_e$ \citep{2009RvMP...81..969H}. However, calculating absolute temperatures is uncertain because of the requisite assumptions about the precise ionic fractions and elemental abundances.

Figure \ref{fig:mean_lines} suggest a flattening in the \nii/\ha index in both the \hii regions and the DIG distributions for $r< 0.6R_{25}$, indicating a potential contribution from the DIG to the observed radial metallicity gradients within galactic planes \citepalias{2017MNRAS.466.3217Z}.

On the other hand, the \sii/\ha line ratio increases radially in both \hii regions and DIG regimes (lower-left panel of Fig. \ref{fig:mean_lines}), with higher values observed in the DIG; from 0.28 to 0.63 for the DIG and from 0.19 to 0.12 for the \hii regions between 0.2R$_{25}$ and 1R$_{25}$, with the \sii/\ha DIG line ratio higher by 0.15 dex, on  average. The shapes of both distributions display comparable patterns, reaching a maximum at 0.75$-$0.8R$_{25}$.

The \oi line is produced by collisions of neutral oxygen with thermal electrons, while its intensity is a measure of the neutral hydrogen content within the DIG. The first ionisation potential of the oxygen is close to that of the hydrogen, and the large H$^{+}$+O$^0\leftrightarrow$ H$^0+$O$^{+}$ charge-exchange cross-section keeps O$^{+}/$O nearly equal to H$^{+}/$H. Thus, the \oi/\ha\, ratio is related to the amount of H$^0$ relative to H$^{+}$, and it is a sensitive probe of the ionisation state of the emitting gas \citep{1998ApJ...494L..99R, 2009RvMP...81..969H}. 
The volume photon emissivity \textit{e} of the O$^1$D $\lambda$6300 transition relative to \ha is related to the hydrogen ionisation ratio $n$(H$^+$)/$n$(H$^0$), with a linear dependence of the gas-phase abundance of oxygen $n$(O)/$n$(H) and a weak dependence that tracks changes in $T_e$ \citep{2009RvMP...81..969H}.
Observations of \oi in the Milky Way and objects at $z \sim 0$ are difficult because of the \oi $\lambda$6300 air glow line, which is of order 100 times brighter than the interstellar line. \citet{2006ApJ...644L..29V} made the first DIG detection of \oi $\lambda$6300 in any non-edge-on spiral other than the Milky Way near the \hii region NGC 604 in M33, with observed \oi/\ha ratios in the range $0.038-0.097$.
In the upper-right panel of Fig. \ref{fig:mean_lines}, we observe the azimuthally averaged values of the \oi/\ha ratio with a significantly steeper slope in the DIG compared to the emission in the \hii regions, increasing from 0.12 to 0.79 between 0.25R$_{25}$ to 0.75R$_{25}$ (-0.9 to -0.1 in log), while the \hii distribution increases from 0.06 to 0.18 (-1.2 to -0.7 in log), which are higher than the values reported by \citet{2006ApJ...644L..29V} for M33, and higher than the values found by \citetalias{2022A&A...659A..26B} with an upper limit $\sim 0.3$. The high \oi/\ha line ratio in the DIG found in the BETIS showcase sample is challenging, and will be matter of study in forthcoming studies.

The radial distribution of the \oiii/\hb line ratio is shown in the lower-right panel of Fig. \ref{fig:mean_lines}. Generally, the variation in this ratio between DIG and \hii regions depends on the specific physical characteristics of the ISM \citepalias{2017MNRAS.466.3217Z}. 
Normal spiral galaxies show an increasing value of the \oiii/\hb line ratio with increasing radius, mainly due to secondary dependence on metallicity, with a high dispersion in the central regions ($\sim 1$ dex for $r < 0.4R_{25}$). Previous studies also show a \oiii/\hb ratio both higher in \hii regions \citep{1997ApJ...483..666G, 1999AJ....118.2775G} and lower than the DIG regions \citep{2001ApJ...551...57C, 2002ApJ...572..823O}.
In our sample, this ratio remains elevated in DIG regions, but the difference between the DIG and \hii distributions is less pronounced compared to the other line ratio distributions. This trend suggests a radial decrease in this ratio; from 1.58 to 1.0 in the DIG (0.2 to 0.0 in log) and from 1.07 to 0.79 in the \hii regions (0.05 to -0.1 in log). Our results indicate that these line ratios are typically higher in the DIG compared with the \hii regions. This trend aligns with what is commonly reported in the existing literature \citep{2009RvMP...81..969H}.

The similarity in the trends observed in the radial distribution of \SBha and line ratios between both DIG and \hii regions, especially in the case of \nii/\ha and \sii/\ha ratios, along with the majority of DIG bins falling within the photoionisation regime on the BPT diagram, suggests that the explanation for DIG behaviour can be attributed to photoionisation processes within the galactic plane, without the need for alternative sources of ionisation.

However, another phenomenon needs to be explained, the low values of \EWha\, found in the DIG, specially in regions with photoionisation regimes not corresponding to SF regions \citepalias{2018MNRAS.474.3727L, 2022A&A...659A..26B}.

\begin{figure*}[ht!]
    \includegraphics[width=\textwidth]{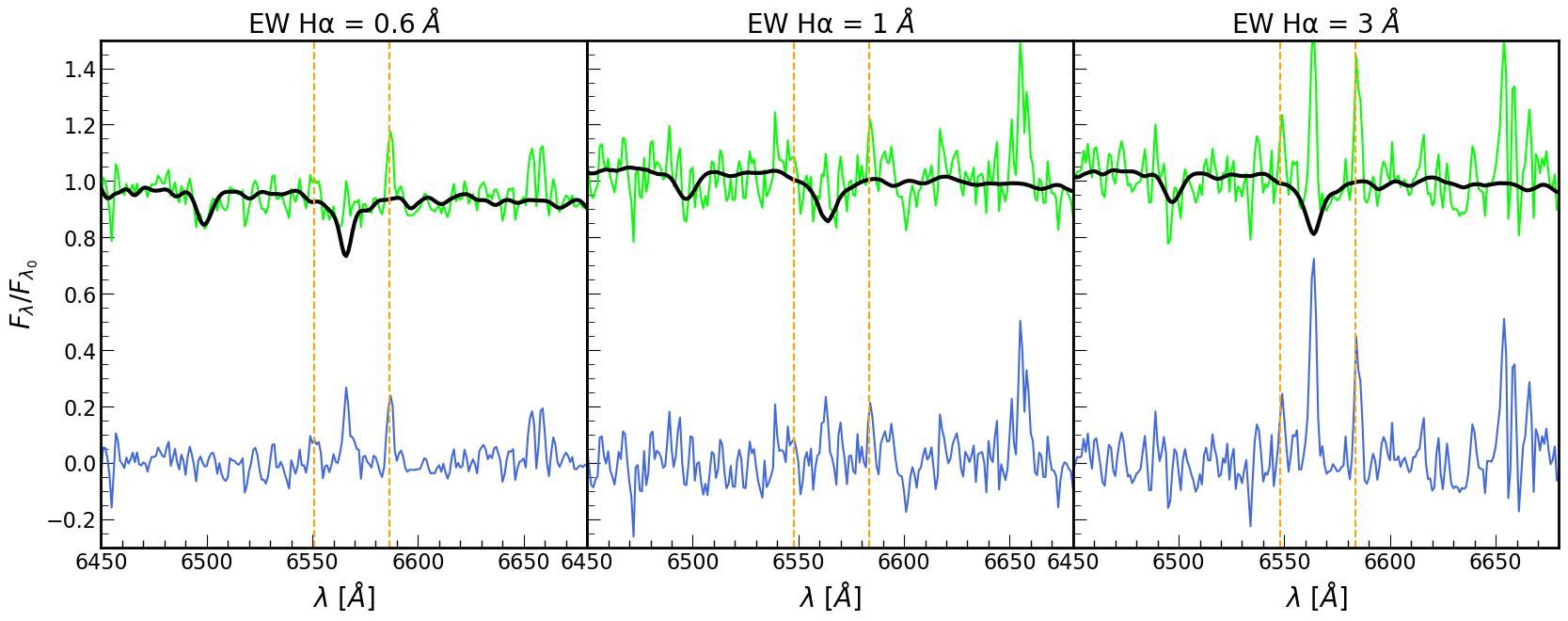}
    \caption{Three examples of observed (green), model (black) and nebular (blue) spectra of different \EWha = 0.6, 1 and 3 \r{A}. All observed spectra are integrated spectra from NGC863 bins, the model spectra are obtained from {\sc STARLIGHT} following our methodology and the nebular spectra are obtained subtracting the model to the observed (see text). Gold lines represent the \nii doublet.}
    \label{fig:EWs_spectra}
\end{figure*}

\begin{figure}[ht!]
    \includegraphics[width=\columnwidth]{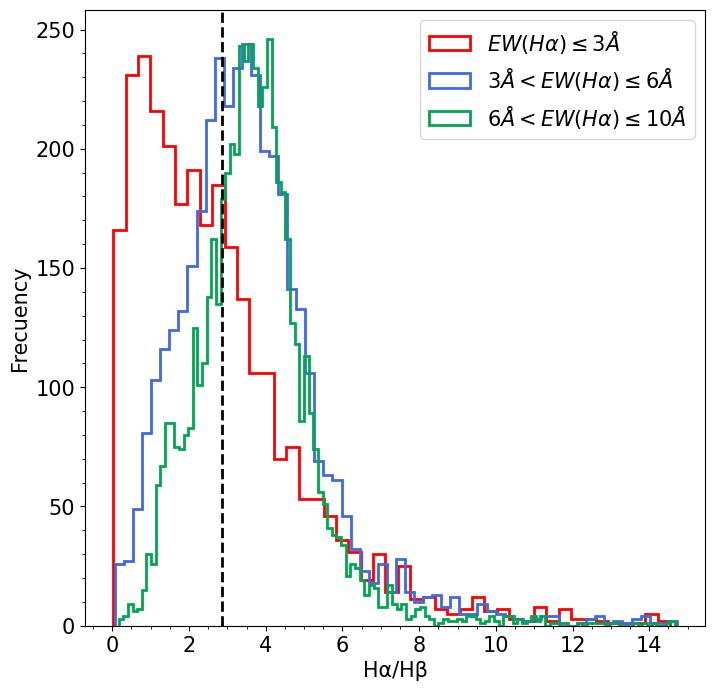}
    \caption{Distribution of \ha/\hb ratio for three different \EWha\, regimes. Each distribution represent the \ha/\hb flux ratio for all bins between 6 \r{A} $<$ \EWha\,$\leq10$ \r{A} (green), 3\r{A}$<$\EWha\,$\leq6$ \r{A} (blue), and \EWha\,$\leq3$ \r{A} (red), for the seven binned galaxies listed in Table \ref{tab:TableSample}. Black vertical dashed line presents the theoretical ratio of \ha/\hb = 2.87 \citep{2006agna.book.....O}. We can see that a substantial part of the bins with \EWha\,$\leq3$ \r{A} shows a non-physical ratio ($<$ 2.87).}
    \label{fig:EWs_Ha_Hb_ratio}
\end{figure}

\subsection{\EWha\, in the DIG}

The \EWha\, has been used by many authors to differentiate between ionisation caused by star formation and AGNs and ionisation caused by a smooth background of hot evolved stars. \citet{2011MNRAS.413.1687C} used SDSS data to demonstrate that the emission-line galaxy population exhibits a bimodal distribution in \EWha, and that 3\r{A} serves as an empirical demarcation between these two. 
Later, \citet{2016MNRAS.461.3111B} using MaNGA data, showed the presence of extended (kpc scale) low-ionisation emission-line regions (LIERs) in both star-forming and quiescent galaxies, associated with low \EWha\,(< 3\r{A}). In SF galaxies, the LIER emission was associated with diffuse ionised gas, most evident as extraplanar emission in edge-on systems.

\citetalias{2018MNRAS.474.3727L} proposed a separation of DIG ionisation regimes based on the \EWha\, and applied over all types of galaxies, including elliptical and S0. The regions where \EWha\,$>14$ \r{A} traces SF regime, $3<$ \EWha$<14$ \r{A} reflects a mixed regime, and regions where \EWha$<3$ \r{A} define the component of the DIG where photoionisation is dominated by hot, low-mass, evolved stars (HOLMES; \citealt{2011MNRAS.415.2182F, 2011MNRAS.413.1687C}). Those stars were proposed as an additional ionisation source of the DIG, in order to explain the high \oiii/\hb ratio found in the extraplanar DIG in edge-on galaxies \citep{1998ApJ...494L..99R, 1999ApJ...521..492R, 2011MNRAS.415.2182F}.

The significance of employing the \EWha to distinguish between various ionisation sources, while considering the contribution of HOLMES to the energy budget, has been a topic of frequent discussion among several authors, even in the context of face-on galaxies (e.g. \citealt{2011MNRAS.413.1687C}; \citetalias{2018MNRAS.474.3727L, 2022A&A...659A..26B}); however it is not yet clear whether it reveals a true division (if any) between ionisation carried by HOLMES, star formation, or shocks. Moreover, \citetalias{2022A&A...659A..26B} also noticed an unclear division for ionisation by HOLMES at \EWha = 3 \r{A} and by star formation at \EWha = 14 \r{A} suggested by \citetalias{2018MNRAS.474.3727L}, since their \EWha\, for DIG and \hii regions tends to overlap in lower \EWha\, regimes. 

However, the general methodology employed in IFS studies in order to measure both the line intensities and the EW of Balmer recombination lines poses a problem for spectra with low S/N values for the emission lines. This methodology calls for measuring the line intensities and EWs from a nebular spectrum obtained by subtracting a fitted stellar model to the observed spectrum. In the case of the Balmer recombination lines, this procedure typically makes a small correction to the emission line intensity due to the underlying stellar absorption. For \EWha $> 3\r{A}$, this correction is usually of a few percent. Nevertheless, when the S/N of the \ha line is $\sim 1$ (\ha emission embedded in the stellar continuum), the resulting \ha emission, and therefore \EWha, would depend on the Balmer stellar absorption feature due to the SSP fitting. In this case, the underlying stellar absorption would not represent just a correction of the emission line, but the resulting flux would be fully originated from the fitted stellar spectrum, generating low \EWha values $\lesssim 3$ \r{A}.

This is shown in Fig. \ref{fig:EWs_spectra}, where we plot the observed (green), fitted stellar model (black) and nebular (blue) spectra of regions with \EWha\, of 0.6, 1, and 3 \r{A}. In the left and central panels, the \ha emission in the observed spectra is totally embedded by the stellar continuum. However, due to the fitted stellar model, the nebular spectra includes a \ha emission line of the same magnitude as the stellar absorption feature of the model, from which an \EWha of 0.6 \r{A} (left) and \EWha = 1 \r{A} (right) are measured. On the other hand, the right panel of Fig. \ref{fig:EWs_spectra} shows that for a region with \EWha = 3 \r{A}, the underlying stellar absorption of the model makes just a small correction to the total flux of the emission line. 

Consequently, when the S/N of the \ha is $\sim 1$ and/or when the emission is embedded in the stellar continuum (\EWha $\lesssim 3$ \r{A}), the resulting \EWha is dependent on the selection of SSPs and the fitting methodology. This sharp differentiation between HOLMES dominated regions, as well as the differentiation between LIERs and passive galaxies \citep{2011MNRAS.413.1687C, 2016MNRAS.461.3111B} may not be reliable, as it could potentially originate from a methodological artefact. We show a test of this effect for an integrated region of NGC863 in Appendix \ref{sec:appen_B}.

\begin{figure*}[ht!]
    \centering
    \includegraphics[width=0.9\textwidth]{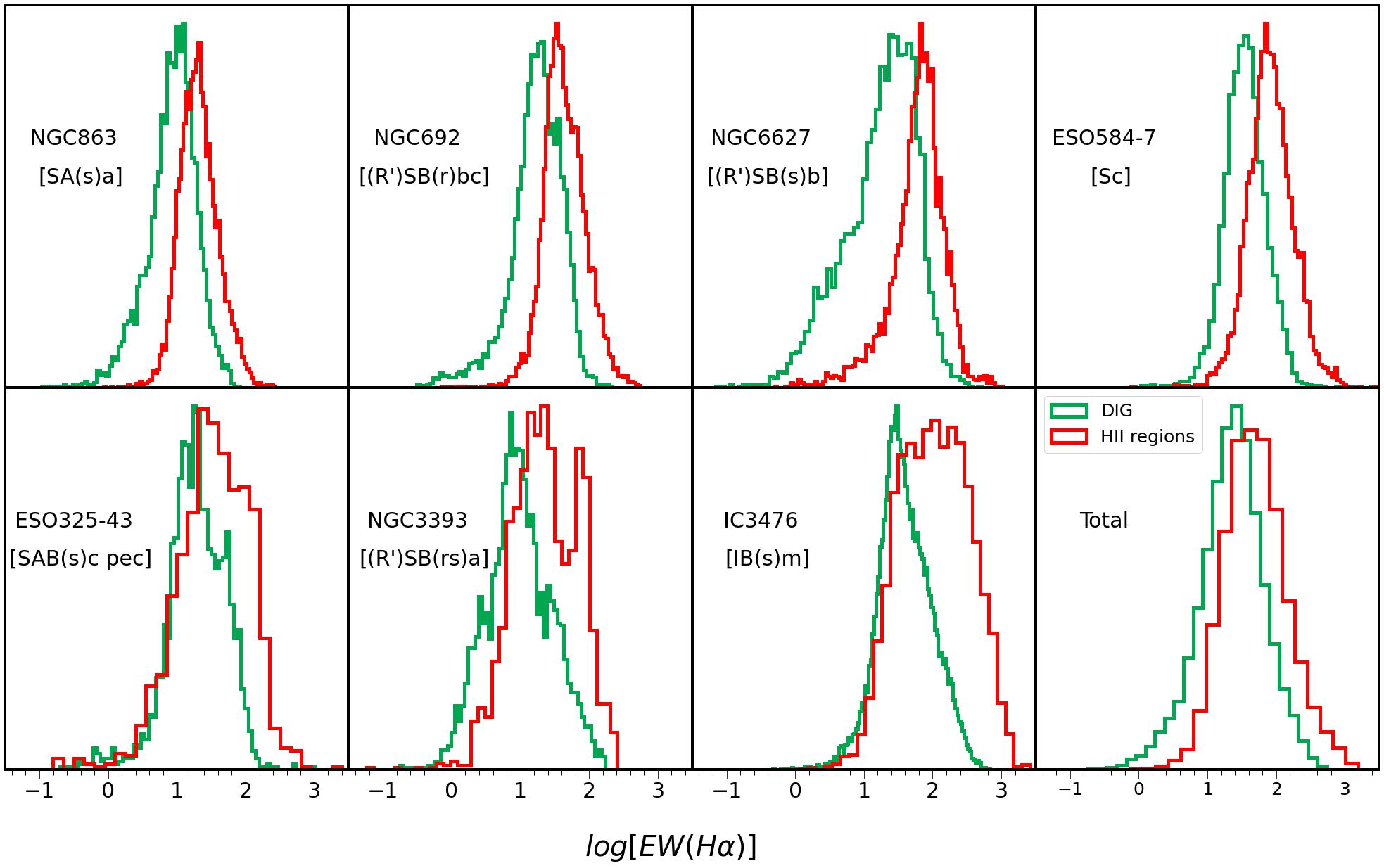}
    \caption{\EWha distributions for all the bins of the sample. Green histograms represents the distribution for the DIG emission bins, and red for the \hii regions emission bins. Last histogram represents the distribution for the seven galaxies.}
    \label{fig:EW_Ha_subsample}
\end{figure*}

Further evidence that the low \EWha regime ($\lesssim 3$ \r{A}) imposes a methodological challenge is manifested when verifying the validity of physical parameters derived from the embedded emission lines, such as the \ha/\hb Balmer decrement. Figure \ref{fig:EWs_Ha_Hb_ratio} shows the distribution of the \ha/\hb ratio for different \EWha regimes, while for \EWha $ >$ 3 \r{A} the distributions peaks at the theoretical \ha/\hb ratio $\gtrsim 2.87$ \citep{2006agna.book.....O}, for the regions with \EWha < 3 \r{A}, the ratio shows non-physical values.

In addition, the values of \EWha< 14 \r{A} and < 3 \r{A} are not exclusive to the DIG emission, so the demarcation with \hii regions using the \EWha is not clear \citepalias{2022A&A...659A..26B}. This is also evident in Fig. \ref{fig:EW_Ha_subsample}, which shows the distribution of \EWha for both DIG and \hii bins within our sample.
We found that the galaxies with lower median \EWha in the DIG regions ($\sim 9$ \r{A}) are NGC863 and NGC3393, both Sa type, followed by NGC692 (Sbc), NGC6627 (Sb) and ESO325-43 (Sc), with \EWha$\sim 18-20$ \r{A}. ESO584-7 (Sc) and IC3476 (Im) have exceptionally high \EWha in the DIG regions, with 34 and 36 \r{A} respectively. ESO584-7 is a \hii galaxy \citep{1998A&AS..130..285C}, thus, the high luminosity and star formation rate may be the cause of this high median \EWha. The case of IC3476 is also special. It is a galaxy suffering the effects of the ram pressure stripping due to the Virgo Cluster intergalactic environment \citep{2021A&A...646A.139B}. These authors show that the effects of this perturbations reach scales of individual \hii regions r$_{eq}\sim 50$ pc. Furthermore, the compression of the gas along its stellar disc may be the cause of the increase of the star formation activity. This increment could explain the exceptionally high \EWha in the DIG regions found in this galaxy, as happened in the case of ESO584-7. 

In general, this results are in concordance with previous authors \citepalias{2018MNRAS.474.3727L, 2022A&A...659A..26B}, being the early type galaxies those with lower \EWha, due to the older stellar populations of their bulges, and late-types those with higher \EWha. The median \EWha of the DIG regions for all sample is $\sim 25$ \r{A}, substantially higher in comparison with previous studies ($\sim5$\r{A}) due to the bias given by ESO584-7 and IC3476, and due to the small sample selected. 

\subsection{Dust reddening in the DIG regime}

\begin{figure}[ht!]
    \includegraphics[width=\columnwidth]{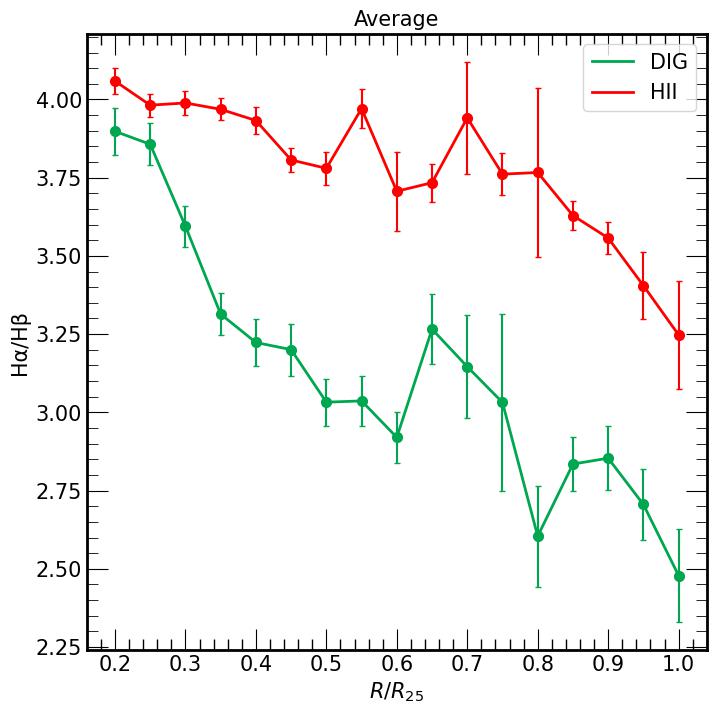}
    \caption{Mean radial \ha/\hb ratio distribution of the sample. The distributions are obtained as in Fig. \ref{fig:mean_lines}. Red represent the distribution for the \hii regions bins, and green for the DIG bins. Both distributions decrease radially, on average, but the extinction is always lower in the DIG regime.}
    \label{fig:Ha_Hb_subsample}
\end{figure}

In Fig. \ref{fig:Ha_Hb_subsample}, we show the reddening in the DIG bins, obtained as the radial distribution of the $H\alpha/H\beta$ ratio for the seven galaxies of our sample. This ratio is always lower in the DIG regime (between 3.85 and 2.5), following the same tendency to decrease radially as in the \hii regions, with ratio between 4 and 3.25. The decline of the \ha/\hb ratio in DIG regimes shown are in concordance to the expectations. Since the \ha/\hb ratio reflects the attenuation of young stars by dust both in \hii regions and in the ISM \citep{2013MNRAS.432.2061C}, this ratio is expected to be lower in the DIG, due to the lower optical depth and the increased scatter in the dust attenuation-line luminosity relation \citep{2020MNRAS.498.4205V}. The fact that we find a \ha/\hb ratio lower in the DIG than in the \hii regions suggests that the extinction is consistently lower in the DIG. There could also be an effect of the temperature, as it has been found in the literature that the DIG temperature is $\sim$ 2000 K warmer than in \hii regions \citep{2005ApJ...630..925M}.

\section{Summary and conclusions}\label{SEC:conclusions}

In this work, we present the Bidimensional Exploration of the warm-Temperature Ionised gaS (BETIS) project, designed for the spatially resolved and spectral study of the diffuse ionised gas (DIG) in a selection of nearby galaxies. 
We present a methodology for characterising and studying, both spatially and spectroscopically, the DIG optimised for galaxies with different linear resolutions and physical characteristics. To validate our methodology, we selected a showcase sample consisting of seven galaxies with diverse morphological and characteristic traits. This methodology involves the following steps:

\begin{itemize}

    \item An adaptive binning was performed to the observed datacube in order to increase the S/N of the fainter lines such as \oiii, \oi, and \sii. Our technique is based on the spectroscopic S/N of the \sii line, with a target S/N of 10.
    
    \item We conducted a SSSP synthesis using the {\sc STARLIGHT} code for each integrated spectrum within the binned datacube. Subsequently, we employed Gaussian fitting to the residuals of each SSP fitting to derive the binned emission line maps for the nine lines of interest. Additionally, we generated a binned \EWha \,map.
    
    \item The DIG was separated from the \hii regions using a combination of an automated tool to detect and subtract the \hii regions from the binned \ha maps, together with a cut-off in \ha surface brightness to subtract bright, irregular \hii regions not detected by the automated tools.
\end{itemize}

We found an average DIG faction of 40\%-70\% in the showcase sample, with NGC3393 the one with the higher DIG fraction (69\%-87\%), followed by NGC863, ESO325-43, NGC692, NGC6627, ESO584-7, and IC3476. Those with higher DIG fractions are the two Seyferts of the sample: NGC3393 and NGC863. This is further exemplified when analysing the radial distributions of the \SBha in the DIG and \hii regions. In these two galaxies, the DIG exhibits higher surface brightness compared to the \hii regions, with the disparity between these two regimes reaching up to 1 dex. The overall radial distributions of \SBha, as depicted in Fig. \ref{fig:SB_radial}, generally reveal similar trends for both \hii regions and DIG, with a radial decrease. However, there is an increase in \SBha within the bars of NGC692 and NGC6627 for the \hii regions. 

On average, we observe in Fig. \ref{fig:mean_lines} higher \sii/\ha, \nii/\ha, \oiii/\hb, and \oi/\ha ratios in the DIG compared to the \hii regions. Additionally, the radial trends of the DIG and \hii distributions are similar in all cases, indicating a correlation between the ionisation of these species in both the DIG and the \hii regions.


Computing the \nii BPT diagram also highlights a significant distinction between the two Seyfert galaxies and the rest of the sample. It is evident that the DIG is predominantly photoionised by \hii regions in all galaxies, except for NGC3393 and NGC863. In these two cases, the ionisation source of the DIG appears to be accounted for by the fast shock models proposed by \citet{2008ApJS..178...20A}. Nevertheless, it is worth noting that these two galaxies host prominent AGNs, which can mimic the emission of the DIG when assessing a global BPT for the entire sample. In particular, NGC3393 presents a strong ionisation cone due to galactic outflows.

We also addressed the challenge of employing the \EWha \,as a proxy for delineating DIG regions and different ionisation regimes. This issue arises because at low \EWha, typically used to identify HOLMES or AGNs regimes, the \ha line derived from synthetic spectra after conducting a SSP fitting can be an artefact of the model. This results in an artificial \ha emission line when correcting the observed \ha emission with the stellar model absorption, after subtracting the model from an observed \ha line with a spectroscopic $S/N\sim1$. Therefore, the low \EWha value may be a result of the characteristics of the stellar models, making it dependent on these models -- if we do not consider the spectroscopic S/N of the observed \ha.

These results suggest that conducting a global analysis of the DIG using a sample of galaxies with diverse characteristics may lead to misleading conclusions about the ionisation mechanisms. This is because each galaxy can present distinct physical processes, for example large-scale AGNs, which can mimic the emission of high-excitation DIG, making the large-scale AGN emission and and high-excitation DIG indistinguishable in the same diagnosis. For this reason, and due to the lack of reliability of the low \EWha\, regimes, every galaxy needs to be considered individually when performing a DIG diagnosis.

The distributions of \EWha for the DIG regions in our sample exhibit a morphological pattern. Sa-type galaxies, with prominent bulges of older stellar populations, have lower \EWha, followed by Sb and Sc galaxies. Notably, ESO584-7 and IC3476 show significantly higher EWs. ESO584-7 is an \hii galaxy, while IC3476 experiences elevated EWs due to ram pressure stripping in the Virgo Cluster's intergalactic environment.

Lastly, we examined the impact of dust reddening in the DIG by assessing the radial distribution of \ha/\hb ratios within both the DIG and \hii regions across our entire sample.Our results suggest that extinction is consistently lower in DIG regions.

In the forthcoming papers of this series, we will apply the methodology and expand the analysis to a full BETIS sample selected from the AMUSING, AMUSING+, and AMUSING++ project samples \citep{2020AJ....159..167L}. We will explore how the results and methodology may vary based on galaxy morphology and inclination (de-projected face-on galaxies $i<45\degree$ vs edge-on galaxies $i>75\degree$), we will investigate the influence of the DIG on the determination of various parameters, including chemical abundances and star formation rates. Additionally, we will explore other spectroscopic lines of interest for the DIG study, such the high-excitation \hei, aiming to uncover potential new ionisation mechanisms of the DIG. Furthermore, we will explore the extragalactic DIG for edge-on galaxies, shedding light on the high-excitation regimes found at at more than 1 kpc above the galactic plane, reported by previous authors.

\begin{acknowledgements}
Based on the observations at ESO with program IDs: 095.D-0172, 098.B-0551, 099.B-0294, 099.D-0022, 0101.D-0748 and 0103.D-0440. R.G.D. acknowledges the CONAHCyT scholarship No. 1088965 and INAOE for the PhD program. The authors also acknowledges Laboratorio Nacional de Supercómputo del Sureste de México (LNS) for allowing the usage of their cluster with the project No. 202201027C in collaboration with INAOE, and Itziar Aretxaga for allowing us the usage of the Toltec/GTM cluster. L.G. acknowledges financial support from the Spanish Ministerio de Ciencia e Innovaci\'on (MCIN), the Agencia Estatal de Investigaci\'on (AEI) 10.13039/501100011033, and the European Social Fund (ESF) "Investing in your future" under the 2019 Ram\'on y Cajal program RYC2019-027683-I and the PID2020-115253GA-I00 HOSTFLOWS project, from Centro Superior de Investigaciones Cient\'ificas (CSIC) under the PIE project 20215AT016, and the program Unidad de Excelencia Mar\'ia de Maeztu CEX2020-001058-M.
The images of Fig \ref{fig:sub_collague} were created with the help of the NOIRLab/IPAC/ESA/STScI/CfA FITS Liberator and the free web-based image editor \href{photopea.com}{photopea.com}.

\end{acknowledgements}

\bibliographystyle{aa}
\bibliography{aa}

\appendix

\section{Emission line measurement} \label{sec:SSP_apendA}

\subsection{Cube preprocessing}

Each individual spectrum of the cube, after being brought to the rest-frame, must be corrected for Milky Way extinction. The correction is carried out by multiplying each spectrum by a factor $10^{0.4\cdot a_{\lambda}}$, where $a_{\lambda}$ is the extinction function of \cite{1999PASP..111...63F}, using the Python library \textit{extinction.fitzpatrick99}\footnote{\href{https://extinction.readthedocs.io/en/latest/index.html}{https://extinction.readthedocs.io/en/latest/index.html}}, that reads the $R_V=A_V/E(B-V)$ ratio (fixed at 3.1) and the total V-band extinction in magnitudes for each galaxy, obtained from Hyperleda. Values can be found in the table of BETIS characteristics online. 

In this work we use CB07 \citep{2007IAUS..241..125B, 2007ASPC..374..303B} base spectra for the SSP models. The model spectra have a resolution of $R\sim 2000$ and $1\lambda$ of spectral sampling. Our observed spectra have $R\sim 3000$ and $1.25\lambda$ of spectral sampling, so the observed spectra must be resampled at $1\lambda$. Performing a linear interpolation of every single spectra of the cubes resolves the conflict between the samplings. After these steps, we obtain a new observed cube, rest-framed, resampled to $\Delta\lambda=1$ \r{A} and corrected to Milky Way extinction. This new resampled cubes are those using to perform the SSP fittings.

\subsection{SSP synthesis}

Performing a SSP synthesis \citep{1968ApJ...151..547T} requires an estimation of the type of stellar populations, namely, the masses, ages, and metallicities found in a galaxy, star cluster, or region of a galaxy based upon its spectra. 
The SSP synthesis is carried by the  {\sc STARLIGHT} software \citep{2005MNRAS.358..363C}, which fits an observed spectrum, $O_{\lambda}$m with a model, $M_{\lambda}$, which adds
up N* spectral components from a pre-defined set of base spectra. The synthetic model spectra that the program generates take the form:

\begin{equation}
    M_{\lambda}=M_{\lambda_0} \left( \sum^{N_*}_{j=1}x_j b_{j,\lambda}r_\lambda \right) \otimes G(v_*, \sigma_*),
        \label{eq:SSPmodels}
\end{equation}

where $b_{j,\lambda}$ is the spectrum of the \textit{j}th SSP normalised; $r_\lambda$ is the reddening term; $M_{\lambda_0}$ is the synthetic flux at the normalisation wavelength; $\otimes$ is the convolution operator; $G(v_*, \sigma_*)$ is a Gaussian
distribution centred at velocity, $v_*$ and with a dispersion, $\sigma_*$; and \textbf{x} is the population vector. Each component $x_j$ $(j = 1, ... , N_*)$ represents the fractional contribution of the SSP with age, $t_j$, and metallicity, $Z_j$, to the model flux.

As previously mentioned, we use the CB07 base spectra for the fittings.
The $N_*=100$ SSPs comprises three metallicities: $Z=0.2, 1,$ and $2.5Z_\odot$, and 15 ages, from $t=0.001$ to $t=13$ Gyr. All SSPs are
normalised to $1M_\odot$ at $t = 0$.
Their spectra were computed with Padova-2004 evolutionary tracks models, and \citet{2003PASP..115..763C} IMF $\left(0.1M_\odot<M<100M\odot\right)$.

The CB07 base comprises SSPs with the same metallicities and age range than the BC03 base \citep{2003MNRAS.344.1000B}. The difference between them are the TP-AGB spectra that CB07 adds in the base and the use of MILES-2007 evolutionary tracks models instead of Padova-2004.
For the preliminary results in this work, we performed the SSP fitting on the CB07 base spectra.

{\sc STARLIGHT} takes as its input: the wavelengths $(\lambda)$, the observed spectrum ($O_\lambda$, resampled, rest-framed and corrected by MW extinction), the errors $(e_\lambda)$, a base-master file with the SSPs, and a mask file. The mask file contains the regions of the spectra that we do not want to model with {\sc STARLIGHT}, such as emission lines, artefacts, and holes in $O_{\lambda}$. Once {\sc STARLIGHT} is running, it goes on to build the synthetic spectra $M_\lambda$ of the form of Eq.~(\ref{eq:SSPmodels}) and to find the one that best fits the observed $O_\lambda$. The outputs files contains the population mixture of the best fit; the $x_j$, $Z_j$, $t_j$, $(L/M)_j$, stellar masses and the percentage of each component, among the synthetic spectrum $M_\lambda$.

\begin{figure}[ht!]
    \centering
    \includegraphics[width=\columnwidth]{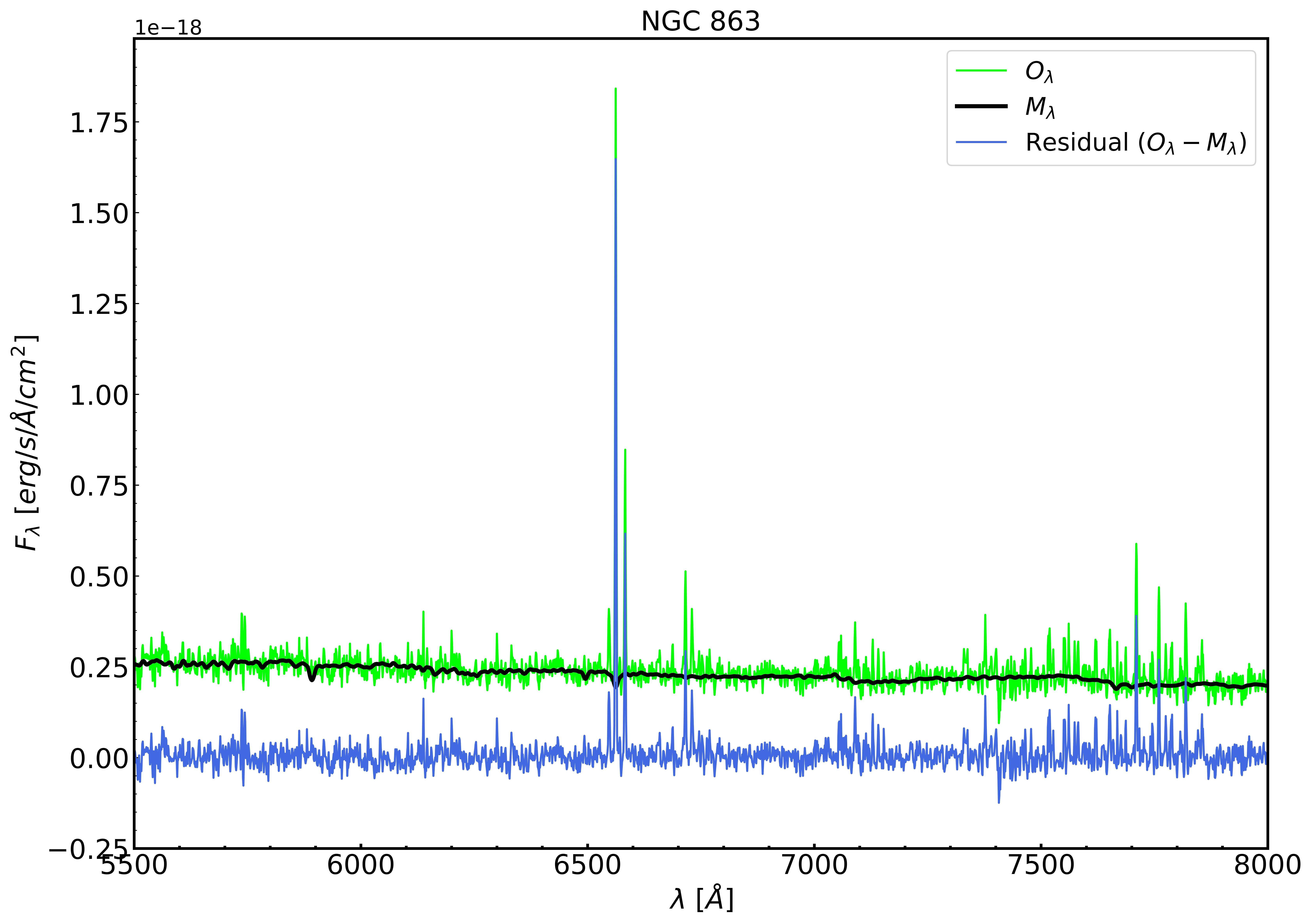}
    \caption{Example of a SSP fitting for a simple spaxel of the galaxy NGC 863. The observed, rest-framed and resampled spectrum is coloured in green. The black spectrum is the best fit synthetic spectrum we get from the {\sc STARLIGHT} SSP fitting, and the blue spectrum is the nebular emission of the spaxel, obtained from the residue of the fit, subtracting the synthetic spectrum to the observed.}
    \label{fig:STARLresidualExample}
\end{figure}

Once we run {\sc STARLIGHT} with all spaxels of our MUSE datacubes ($\sim 100.000$ spectra per cube), we can build a 'nebular gas' or 'residue' datacube of the form $O_\lambda-M_\lambda$, containing the nebular emission of the object (see Fig. \ref{fig:STARLresidualExample}).

\begin{figure}[ht!]
    \centering
    \includegraphics[width=\columnwidth]{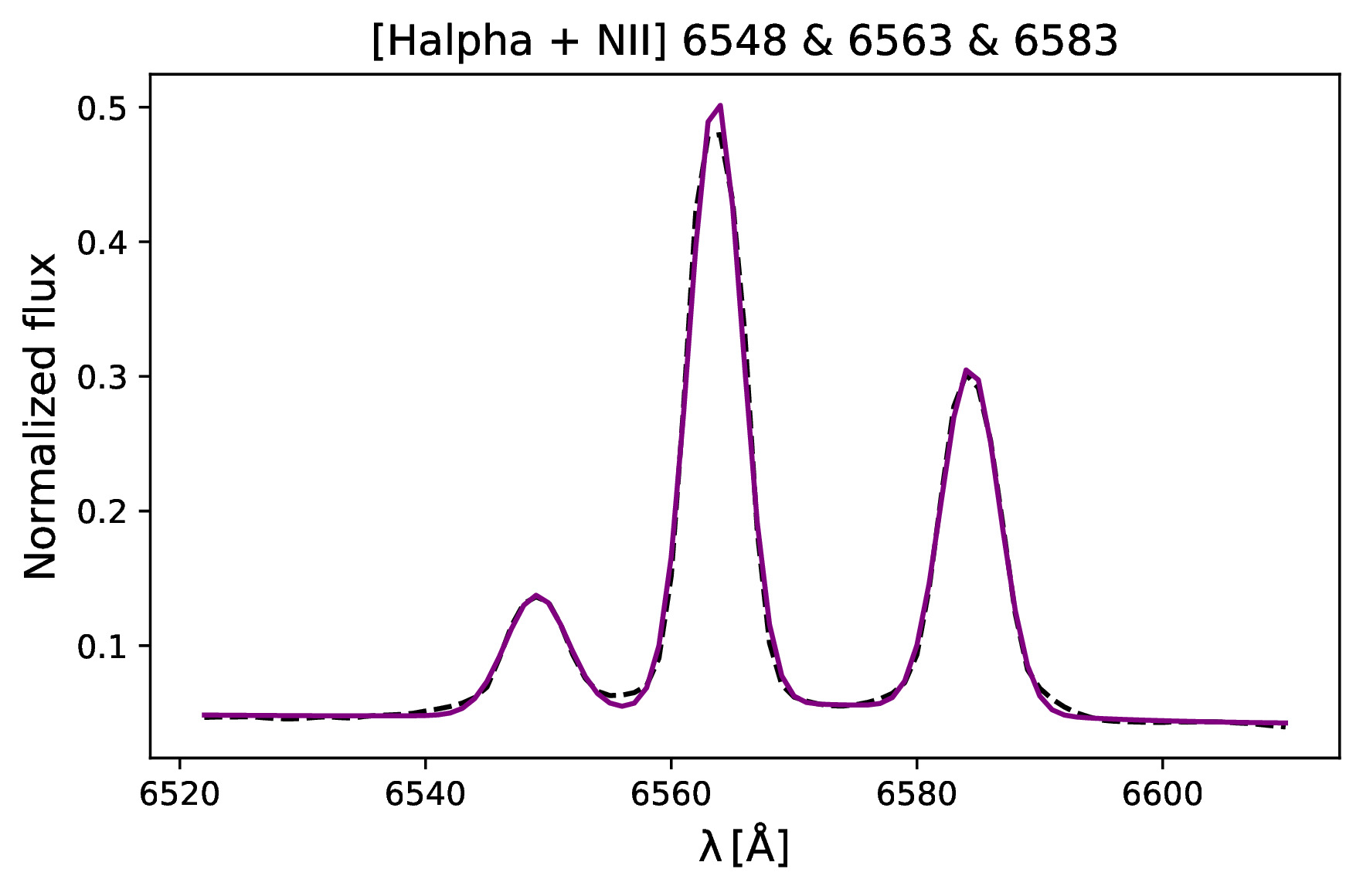}
    \includegraphics[width=\columnwidth]{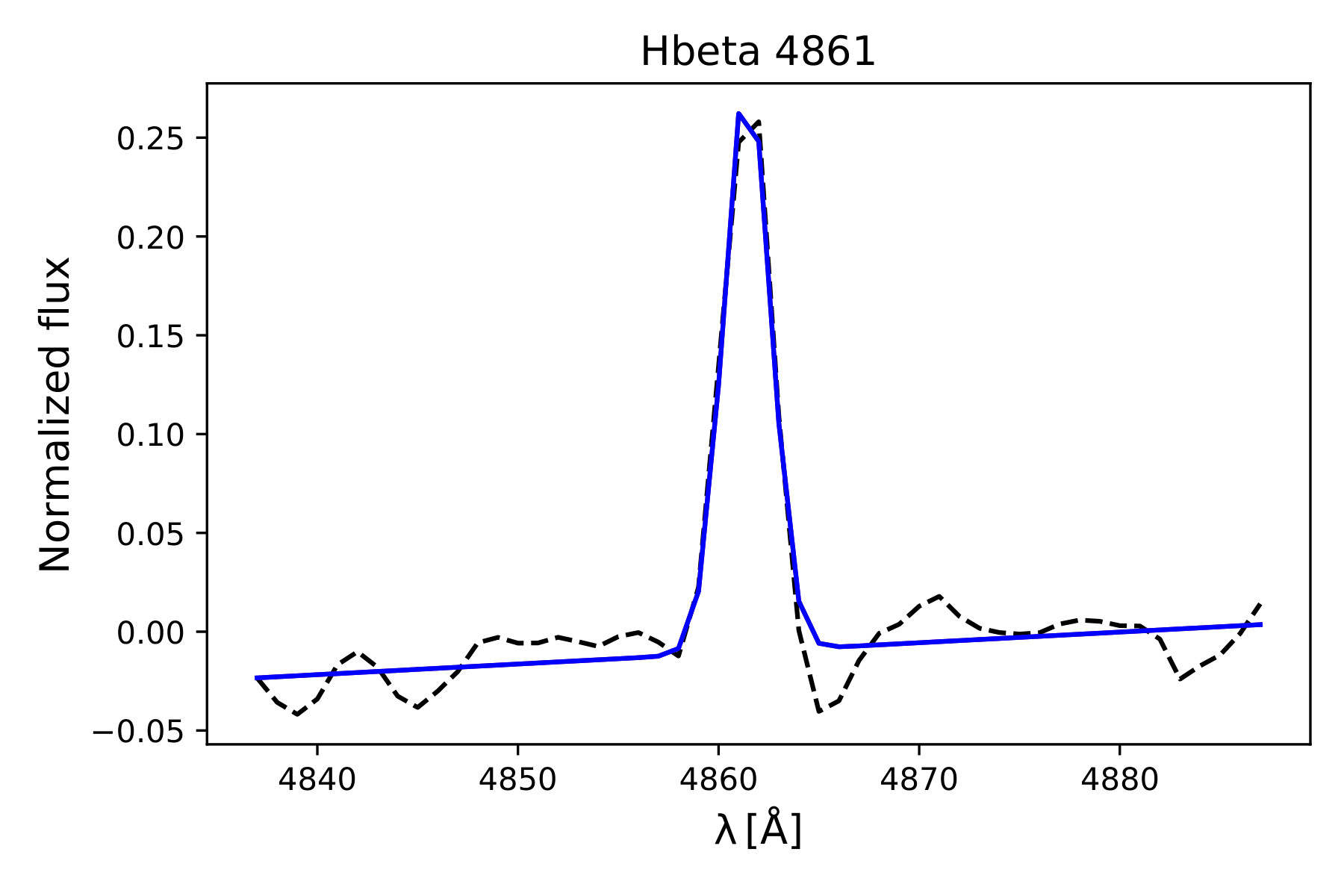}
    \caption{Example of the multipeak fitting of \ha\, and \nii lines (top). The purple curve is the Gaussian curve fitted.
    Example for $H\beta$ (bottom). Blue curve is the Gaussian curve fitted.
    In both cases, black dashed line is the input spaxel spectrum.}
    \label{fig:FITexample}
\end{figure}

Then, using the Python {\sc MPFIT}\footnote{\href{https://github.com/segasai/astrolibpy/blob/master//MPFIT/MPFIT.py}{https://github.com/segasai/astrolibpy/blob/master/MPFIT/MPFIT.py}} module, we perform a Gaussian fitting around the emission lines of interest for all the spaxels of the nebular gas cube. 
This module performs a multipeak Gaussian fit for each individual line, with $\sigma=FWHM_i/(2\sqrt{2ln2})$, with $FWHM_i$ as the MUSE initial full width at half maximum, corresponding to $3$ \r{A}. The module search the peak of the line in a $10$ \r{A} (200-300 km/s) range centred in a central wavelength $\lambda_c$ and performs the Gaussian fit in a range between $0.5\sigma-2\sigma$. The flux of the $i,j$ spaxel for the $\lambda$ line is then defined as $f_{i,j}(\lambda)=\sqrt{2\pi} I_{peak}$, and $I_{peak}$ the peak of the Gaussian fit (see Fig. \ref{fig:FITexample}). 

\section{\EWha\, for different SSPs}\label{sec:appen_B}

\begin{figure}[ht!]
    \includegraphics[width=\columnwidth]{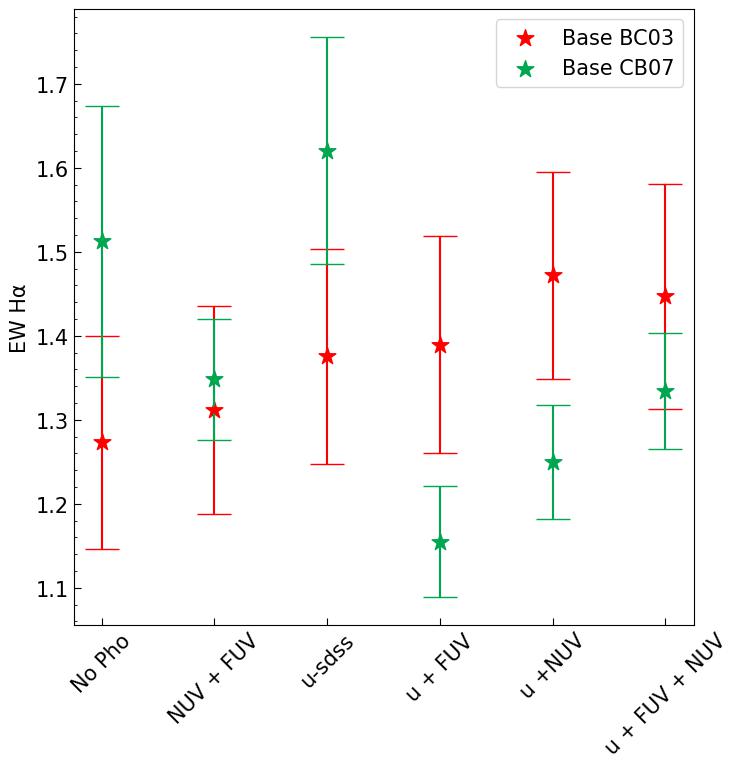}
    \caption{EW($H\alpha)$ measurement for a NGC863 bin of EW($H\alpha)\approx 1$ \r{A} using 12 different SSP fittings with the  \textit{Photometric {\sc STARLIGHT}} code \citep{2019MNRAS.483.2382W}. Red values correspond to the those obtained from the synthetic spectrum after setting the BC03 base. Idem for the green, using the CB07 base. The x-axis represent the six different photometric constraints; 1) No photometric constraints; 2) GALEX-NUV and GALEX-FUV photometric points added to the fitting; 3) only u-SDSS point added; 4) u-SDSS and GALEX-FUV points added; 5) u-SDSS and GALEX-NUV points added; 6) u-SDSS, GALEX-FUV, and GALEX-NUV points added to the fitting. We can see that the type of fitting affects the EW($H\alpha)$ measurement by as much as $80\%$.}
    \label{fig:models_EWs}
\end{figure}

Figure \ref{fig:models_EWs} shows the different values of EW($H\alpha)$ obtained in function of the type of SSP fitting. We used an alternative version of {\sc STARLIGHT}, called {\sc photometricSTARLIGHT} \citep{2019MNRAS.483.2382W}, that combines spectroscopic and photometric constraints to perform SSP synthesis using photometric points to extrapolate the model spectra to the bluer part, solving the problem of the lack of blue constraint in the MUSE spectra due to its spectral coverage. We used the spectrum of a bin with EW($H\alpha) \approx 1$ \r{A} of the NGC863 observed cube binned performing the methodology previously exposed. We selected three different photometric constraints to perform this fittings;  near-ultraviolet (NUV, $\lambda_{eff}=2310$ \r{A}) and far-ultraviolet (FUV, $\lambda_{eff}=1528$ \r{A}) from All-Sky Survey of the Galaxy Evolution Explorer (AIS-GALEX; \citealt{2017ApJS..230...24B}) and u-SDSS ($\lambda_{eff}=3543$ \r{A}) from Sloan Digital Sky Survey (SDSS) DR16 \citep{2020ApJS..249....3A}. Then, {\sc photometricSTARLIGHT} reads the spectrum and the AB magnitudes obtained from the GALEX and SDSS images and performs the SSP fitting setting a certain base spectra (see Sect. 3.2). We make this fittings selecting both BC03 and CB07 bases and selecting all different combinations of photometric constraints with the three photometric points, obtaining then 12 different model spectra. Considering this effect as a cause of systematic error, we can get an error as high as $80\%$ due to the selection of stellar populations considered in the fitting and the type of SSP fitting performed, at low EW($H\alpha)$ regimes.  



\label{lastpage}

\end{document}